%% file: paper3_rev_clean.tex
\documentclass[twocolumn]{aastex62}
\usepackage{CJK}
\usepackage{natbib}
\newcommand{\sersic}{S\'{e}rsic}
\newcommand{\galfit}{\texttt{GALFIT}}

\makeatletter

\newcommand{\Rmnum}[1]{\expandafter\@slowromancap\romannumeral #1@}
\makeatother

\begin{document}
\begin{CJK*}{UTF8}{gkai}
\title{The Carnegie-Irvine Galaxy Survey. \Rmnum{8}. Demographics of Bulges along the Hubble Sequence}

\author[0000-0003-1015-5367]{Hua Gao (高桦)}
\affiliation{Department of Astronomy, School of Physics, Peking University, Beijing 100871, China}
\affiliation{Kavli Institute for Astronomy and Astrophysics, Peking University, Beijing 100871, China}

\author[0000-0001-6947-5846]{Luis C. Ho}
\affiliation{Kavli Institute for Astronomy and Astrophysics, Peking University, Beijing 100871, China}
\affiliation{Department of Astronomy, School of Physics, Peking University, Beijing 100871, China}

\author[0000-0002-3026-0562]{Aaron J. Barth}
\affiliation{Department of Physics and Astronomy, University of California at Irvine, 4129 Frederick Reines Hall, Irvine, CA 92697-4575, USA}

\author[0000-0001-5017-7021]{Zhao-Yu Li}
\affiliation{Department of Astronomy, Shanghai Jiao Tong University, Shanghai 200240, China}

\begin{abstract}
  We present multi-component decomposition of high-quality $R$-band images of
  320 disk galaxies from the Carnegie-Irvine Galaxy Survey. In addition to
  bulges and disks, we successfully model nuclei, bars, disk breaks,
  nuclear/inner lenses, and inner rings. Our modeling strategy treats nuclear
  rings and nuclear bars as part of the bulge component, while other features
  such as spiral arms, outer lenses, and outer rings are omitted from the fits
  because they are not crucial for accurate bulge measurements. The error budget
  of bulge parameters includes the uncertainties from sky level measurements and
  model assumptions. Comparison with multi-component decomposition from the
  \textit{Spitzer} Survey of Stellar Structure in Galaxies reveals broad
  agreement for the majority of the overlapping galaxies, but for a considerable
  fraction of galaxies there are significant differences in bulge parameters
  caused by different strategies in model construction.  We confirm that on
  average bulge prominence decreases from early to late-type disk galaxies,
  although the large scatter of bulge-to-total ratios in each morphological bin
  limits the application of Hubble type as an accurate predictor of
  bulge-to-total ratio. In contrast with previous studies claiming that barred
  galaxies host weaker bulges, we find that barred and unbarred spiral galaxies
  have similar bulge prominence.
\end{abstract}


\section{Introduction}
\label{sec:introduction}

Motivated by the wealth of information stored in the morphological structures of
galaxies, \citet{2011ApJS+Ho} initiated the Carnegie-Irvine Galaxy Survey (CGS)
to investigate the optical photometric properties of 605 bright galaxies in the
southern hemisphere. Detailed analysis of the high-quality CGS images has
yielded significant insights into many aspects of the Hubble sequence of
galaxies, including the nature of disk breaks \citep{2011ApJS+Li}, the formation
history of ellipticals \citep{2013ApJ+Huang1,2013ApJ+Huang2,2016ApJ+Huang}, the
bar buckling phenomenon \citep{2017ApJ+Li}, the nature of S0s
\citep{2018ApJ+Gao}, and the origin of spiral arms \citep{2018ApJ+Yu2,
  2018arXiv+Yu}. Galaxy bulges are, of course, one of the fundamental, defining
characteristics of the Hubble sequence, and hence constitute a central theme of
this long-term program.

As ellipticals and bulges bear resemblance in many aspects of their
observational properties \citep[e.g.,][]{1977egsp.conf+Faber,1977ARA&A+Gott,
  1999fgb+Renzini}, they were once thought to have similar origin: both form out
of rapid, violent processes, such as gravitational collapse \citep{1962ApJ+ELS}
and mergers \citep{1977egsp+Tmoore,2016ASSL+Bournaud}. However, in recent
decades, there has been increasing appreciation that bulges actually constitute
a heterogeneous population. Bulges in late-type spirals show a younger stellar
population, more flattened stellar light distribution, and more
rotation-dominated kinematics compared with bulges in early-type disks
\citep[e.g.,][]{1997ARA&A+Wyse,2004ARA&A+Kormendy,2008AJ+Fisher,2009ApJ+Fisher,
  2016ASSL+Laurikainen}. These dichotomies in bulge properties suggest distinct
formation physics. In addition to violent processes, secular evolution,
facilitated by nonaxisymmetries in the galaxy potential, is able to transport
gas with low angular momentum to galaxy centers to build up bulges that
resembles disks rather than merger-built ellipticals
\citep[e.g.,][]{2004ARA&A+Kormendy,2005MNRAS+Athanassoula,2014RvMP+Sellwood,
  2016MNRAS+Tonini}. The disky bulges formed in this manner are commonly
referred to in the literature as pseudobulges, to distinguish them from
classical bulges.

Despite the importance of bulges in defining the Hubble sequence and their rich
formation physics, robust quantitative measurements of their structural
parameters are yet to be achieved for large, well-defined samples.
One-dimensional (1D) fitting of galaxy surface brightness profiles
\citep[e.g.,][]{1977ApJ+Kormendy1,1977ApJ+Kormendy2,1979ApJ+Burstein} and
two-dimensional (2D) fitting of galaxy images \citep[e.g.,][]{1989MNRAS+Shaw,
  1995ApJ+Byun,1996A&A+de_Jong2} are the two widely employed parametric
techniques. Non-parametric techniques are less often used because of the
difficulty of applying them to nearly face-on cases \citep[e.g.,][]{1986AJ+Kent,
  1987AJ+Capaccioli,1990A&A+Scorza,1990A&A+Simien}. In terms of parametric
fitting, 2D methods are superior because they preserve the maximum amount of
spatial information on morpholically distinct components and because they
conserve flux during the convolution process (\citealp{1995ApJ+Byun,
  1996A&A+de_Jong2}; see Section~1 of \citealp{2017ApJ+Gao} for a review of the
methods). Nevertheless, both 1D and 2D methods suffer from the uncertainties
introduced by the non-uniqueness of input surface brightness models.

In order to clarify which morphological features are most essential in 2D model
construction when the primary intent is to obtain robust bulge parameters,
\citet{2017ApJ+Gao} selected a representative sample from CGS that covers a
sufficiently wide range of morphological features (bars, lenses, rings, and
spiral arms) and explored the impact of modeling the secondary morphological
features. They showed that modeling nuclear and inner lenses/rings and disk
breaks has considerable impact on bulge parameters, whereas outer lenses/rings
and spiral arms have a negligible effect. For example, failure to model disk
breaks or lenses introduces errors that can be as large as $\sim50\%$ in
bulge-to-total ratio ($B/T$) for barred galaxies (see also
\citealp{2014ApJ+Kim}). This important effect is generally ignored in many
decomposition studies (e.g., \citealp{2011ApJS+Simard,2015MNRAS+Meert,
  2015ApJS+Salo}; but see \citealp{2005MNRAS+Laurikainen,2016MNRAS+Kim,
  2017A&A+Mendez-Abreu}). These alarming uncertainties compel us to measure a
new set of bulge parameters for nearby galaxies, derived in a consistent manner
following the optimal strategy defined in \citet{2017ApJ+Gao}.

Toward this end, \citet{2018ApJ+Gao} successfully decomposed 62 CGS S0 galaxies.
This paper extends our previous work and presents a comprehensive catalog of
bulge parameters for 320 non-edge-on disk galaxies in CGS. Definition of the
sample and description of the data are given in Section~\ref{sec:sample-data}.
We closely follow and expand the strategy in \citet{2017ApJ+Gao} to decompose
the galaxies, as detailed in Section~\ref{sec:methodology}. We compare the CGS
bulge parameters with those from the \textit{Spitzer} Survey of Stellar
Structure in Galaxies (S$^{4}$G; \citealp{2010PASP+Sheth}) in
Section~\ref{sec:comparison-with-s4g}. In Section~\ref{sec:bulge-prom-along}, we
study how $B/T$ is distributed along the Hubble Sequence.
Section~\ref{sec:summary} summarizes the paper.

\section{Sample and Data}
\label{sec:sample-data}

The CGS sample is defined by $B_{T}\leq12.9\,\mathrm{mag}$ and
$\delta<0\arcdeg$, without any reference to morphology, size, or environment.
Details of the observations and data reduction are given in \citet{2011ApJS+Ho}
and \citet{2011ApJS+Li}, and will not be repeated here. We focus on the images
taken in the $R$ band because this filter is less sensitive to dust
extinction and young stars; we avoid the $I$ band because its point-spread
function (PSF) suffers from the ``red halo'' effect (see Appendix~A of
\citealp{2013ApJ+Huang1}). The majority of the $R$-band images are of high
quality: the median seeing is $1\farcs01$, the median surface brightness depth
is 26.4\,mag~arcsec$^{-2}$, and the field-of-view of $8\farcm9\times8\farcm9$ is
large enough to ensure robust sky determination for most of the galaxies.

The sample analyzed in this paper is an extension to the sample of S0s presented
in \citet{2018ApJ+Gao}. We add 304 non-edge-on spirals selected with
morphological type index $0<T\leq9.5$ and inclination angle $i\leq70\arcdeg$.
During the course of performing the decomposition, we ended up removing 46
galaxies for a variety of reasons: two galaxies do not have $R$-band images; one
galaxy is edge-on with a razor-thin disk; two galaxies do not have photometric
calibration; two galaxies are highly dust-attenuated and are located in fields
extremely crowded with stars; ten galaxies are highly irregular; three galaxies
do not yield reasonable fits; and 26 galaxies are bulgeless. This leaves 258
spirals with measurable bulges, which, when combined with the 62 S0s, results in
a final sample of 320 disk galaxies. The bulgeless galaxies are of particular
interest \citep[e.g.,][]{2010ApJ+Kormendy2, 2011ApJ+Fisher}, but they are beyond
the scope of this paper. We list them in Appendix~\ref{sec:bulgeless-galaxies}
for future consideration. A brief description of the main morphological features
of each successfully decomposed galaxy is given in
Appendix~\ref{sec:notes-indiv-galax}.

Figure~\ref{fig:sample} summarizes the basic parameters of the successfully
decomposed sample of 320 galaxies. The stellar masses ($M_{\star}$) for 313
galaxies were derived from the total $K_s$ magnitudes compiled in
\citet{2011ApJS+Ho} and mass-to-light ratios following Equation~(9) in
\citet{2013ARA&A+Kormendy}:
\begin{equation}
  \label{eq:ML}
  \log\, (M/L)_{K_{s}}=1.055(B-V)-0.9402,
\end{equation}
using ($B-V$) colors from CGS \citep{2011ApJS+Li}. The galaxies in the sample
are nearby (median $D_{L}=26.1\,\mathrm{Mpc}$) and massive [median
$\log\, \left(M_{\star}/M_{\sun}\right)=10.57$]. Note that some of the S0s have
morphological type indices $T<-3$ due to the inclusion of misclassified
ellipticals \citep{2018ApJ+Gao}. The sample is dominated by intermediate-type
spirals (Sb--Scd). When divided into three subsamples according to the presence
of a bar or lens, we find that the lens galaxies are exclusively S0s and
early-type spirals. They are, on average, the reddest and most massive galaxies
in the sample. The barred galaxy sample is offset to slightly earlier Hubble 
types, redder colors, and higher masses compared with the unbarred objects.
 
\begin{figure*}
  \epsscale{1.15}
  \plotone{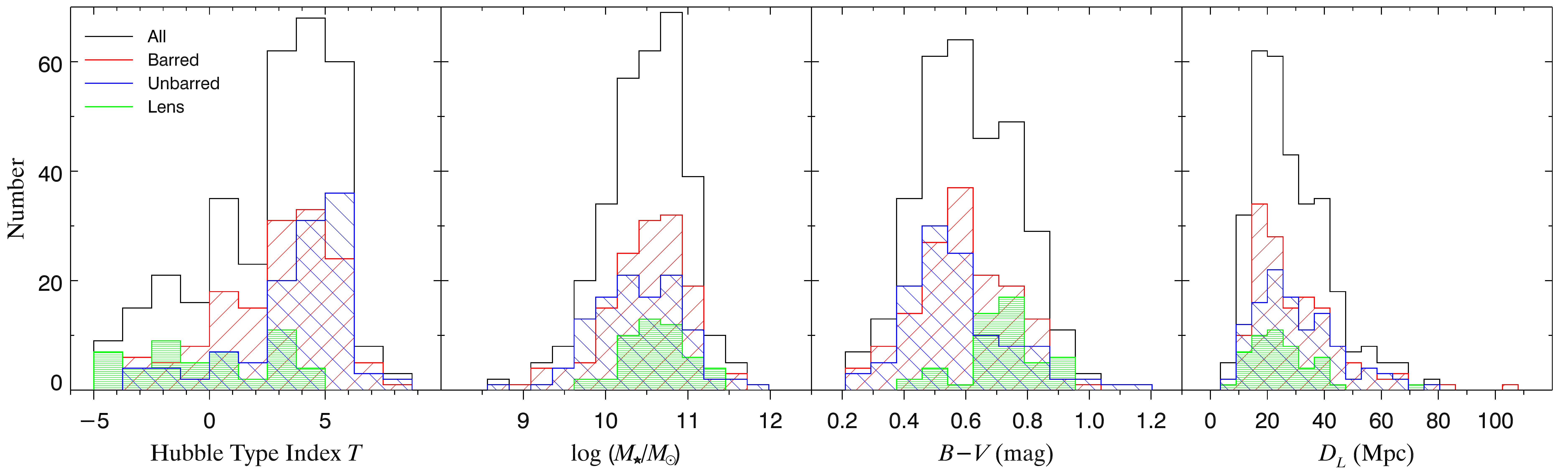}
  \caption{Basic properties of the sample. From left to right, distribution of
    morphological type index $T$, stellar mass, optical color \bv{}, and
    luminosity distance ($D_{L}$). The solid histograms represent the
    successfully decomposed galaxies that are divided into three subsamples:
    barred (red), unbarred (blue), and unbarred galaxies but have a lens
    (hereafter lens galaxies; green). Data from \citet{2011ApJS+Ho} and
    \citet{2011ApJS+Li}. \label{fig:sample}}
\end{figure*}

\section{Methodology}
\label{sec:methodology}

\subsection{Preparation for Image Fitting}
\label{sec:prep-image-fitt}

Following \citet{2017ApJ+Gao}, we use the latest version (3.0.5) of
\galfit{}\footnote{\url{https://users.obs.carnegiescience.edu/peng/work/galfit/galfit.html}}
\citep{2002AJ+Peng,2010AJ+Peng} to perform 2D multi-component decompositions of
the CGS disk galaxies. \galfit{} is a highly flexible and fast image-fitting
algorithm that uses the Levenberg-Marquardt technique to find the optimum
solution. It provides many analytic functions to represent the radial surface
brightness profiles of objects/components of interest, including the widely used
\citet{1968adga+Sersic}, exponential, and modified Ferrer profile. The profiles
are projected onto the image plane via ellipses, or more complicated azimuthal
functions such as Fourier modes, coordinate rotation, and bending modes to break
from axisymmetry, for the purpose of producing realistic-looking galaxies.
Furthermore, each component can be truncated at a given radius and at a given
rate. In practice, however, we only make use of a limited set of its features,
as described in Section~\ref{sec:model-construction}.

For each run, \galfit{} requires a data image, a PSF image, a mask image, and an
input surface brightness model of the galaxy. The data images, PSF images, and
mask images were prepared in \citet{2011ApJS+Ho}. We do not subtract the sky
from the images before fitting, because we aim to solve the sky level
simultaneously during the fit, which has proven to be feasible (see Appendix~B.2
of \citealp{2017ApJ+Gao}). We modify the mask image to account for central dust
lanes that are prevalent in late-type spirals (see notes in
Appendix~\ref{sec:notes-indiv-galax}). Identification of the major dust lanes is
based on $B-R$ color maps. We allow sigma (noise) images to be internally
generated by \galfit{}. The convolution box diameter is set to 40--80 times the
seeing disk, as suggested on the \galfit{} website\footnote{
  \url{https://users.obs.carnegiescience.edu/peng/work/galfit/TFAQ.html}}.  As
mentioned in the Introduction, the uncertainties in bulge parameters introduced
by non-unique models are significant, especially for CGS galaxies that are so
bright and well-resolved that the effects of signal-to-noise ratio and
resolution are marginal. Therefore, an adequate input model is crucial for
deriving accurate bulge parameters. The strategy of preparing models will be
detailed in Section~\ref{sec:model-construction}. 

As the CGS galaxies are bright and well-resolved, the best-fit parameters do not
depend sensitively on their inputs, as long as we provide reasonable initial
guesses. The only exception are the break radius and softening length when
fitting broken disks, but these do not carry much physical significance, and
they are often fixed to reasonable values. Some of the initial guesses are
obtained from \citet{2011ApJS+Ho}, including galaxy centroid, disk ellipticity,
and disk position angle. Other initial parameters, such as surface brightness,
profile shape, ellipticity, and size of bulge, bar, and lens, are estimated
through manual inspection of the image and its isophotal analysis. We identify
the radial range where the component dominates the total light and visually
examine the image and the profiles of surface brightness, ellipticity, and
position angle to estimate their corresponding initial parameters. The initial
guess for break radius is estimated by manual inspection of the surface
brightness profile.

\subsection{Model Construction}
\label{sec:model-construction}

\begin{figure*}[h]
  \epsscale{1.10}
  \plotone{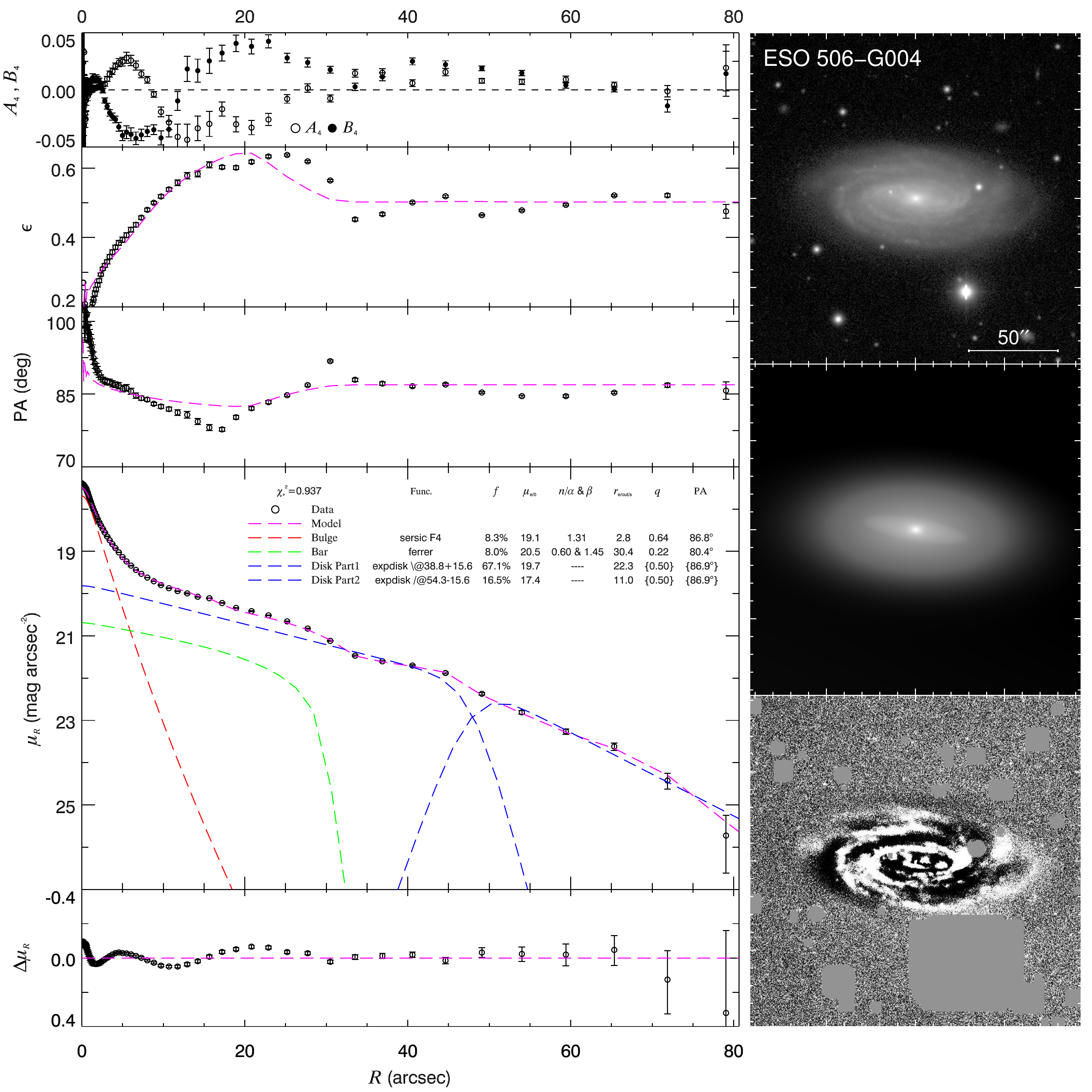}
  \caption{Best-fit model of ESO~506-G004. The left panels display the isophotal
    analysis of the 2D image fitting. From top to bottom, the panels show the
    radial profiles of the fourth harmonic deviations from an ellipse ($A_{4}$
    and $B_{4}$), ellipticity ($\epsilon$), position angle (PA), $R$-band
    surface brightness ($\mu_{R}$), and fitting residuals
    ($\bigtriangleup\mu_{R}$).  Profiles of the data, the model, and the
    individual components are encoded consistently with different symbols, line
    styles, and colors, as explained in the legends. The text to the right of
    the legends gives detailed information on each component; from left to
    right, the columns describe the radial profile functions (PSF, \sersic{},
    exponential, and modified Ferrer) and whether they are complete or truncated
    (blank for complete,
    ``\textbackslash@$r_{\mathrm{break}} +\bigtriangleup r_{\mathrm{soft}}$''
    for outer truncation, and
    ``/@$r_{\mathrm{break}}-\bigtriangleup r_{\mathrm{soft}}$'' for inner
    truncation) and their azimuthal shapes (blank for pure ellipse and F$n$ for
    Fourier modes $m=n$), the light fractions, the characteristic surface
    brightness (effective surface brightness $\mu_{e}$ for the bulge and central
    surface brightness $\mu_{0}$ for the others), the shape parameters of the
    radial profiles (\sersic{} index $n$ for the \sersic{} function and
    $\alpha\,\&\,\beta$ for the modified Ferrer function), the characteristic
    radii (effective radius $r_{e}$ for the \sersic{} function, outer boundary
    $r_{\mathrm{out}}$ for the modified Ferrer function, and scale length
    $r_{s}$ for the exponential function), the axis ratios ($q$), and the
    position angles (PA). The parameters can be constrained to be the same
    (braces) and/or fixed (brackets). Note that the surface brightness profile
    of the model is generated by fixing the geometric parameters to those of the
    data surface brightness profile, and the surface brightness profiles of
    individual components are generated along their major axes; hence, the model
    surface brightness profile is not a simple summation of the profiles of the
    individual components. The right panels display, from top to bottom, the
    grayscale $R$-band image, the best-fit model image, and the residual
    image. The images are shown using the same logarithmic stretch for the data
    and model image, and histogram equalization stretch for the residual
    image. All images are cropped to have the same size of $1.5D_{25}$, with
    $D_{25}$ the isophotal galaxy diameter at $\mu_{B}=25$~mag~arcsec$^{-2}$,
    and are centered on the galaxy centroid, with north up and east to the
    left. \\(The complete figure set for 320 galaxies is available in the online
    journal.) \label{fig:ESO506-G004}}
\end{figure*}

Based on the lessons learned from our detailed study of 2D decomposition methods
\citep{2017ApJ+Gao}, we are aware of which parts of the galaxy should be modeled
or can be omitted, and of the penalties for ignoring certain parts of the galaxy
in the model. Therefore, we prepare just a \textit{single} model for each
galaxy, based on identification of its morphological features through detailed
inspection of the images, color maps, structure maps, and isophotal analysis
products from \citet{2011ApJS+Ho} and \citet{2011ApJS+Li}. Specifically, we
recognize a bulge as extra light above the inward extrapolation of the disk. We
identify a strong bar according to its peak in the ellipticity profile and its
roughly constant position angles at the radial range where its light dominates;
the images, color maps, and structure maps provide extremely useful additional
diagnostics when a bar is weak or viewed end-on.  A lens can be recognized as a
shelf on the surface brightness profile, featuring a sharp decline in surface
brightness in its outskirts.  Spiral arms and rings are readily identified by
visual examination of the images, color maps, and structure maps. In addition to
bulges and disks, we model bars, disk breaks, nuclear and inner lenses, and
inner rings, but do not treat nuclear rings or bars separately because we
consider them as part of the bulge. As with \citet{2018ApJ+Gao}, we model
nuclear point sources (hereinafter nuclei) when present; they can be identified
as abrupt changes in the central color profile.  Spiral arms and outer
lenses/rings have been shown to be not crucial for measuring accurate bulge
parameters, and thus will not be treated.  Unless specifically noted in
Appendix~\ref{sec:notes-indiv-galax}, we follow the above rules to construct
surface brightness models for all the galaxies. The best-fit models of 320 CGS
disk galaxies are shown in Figure~\ref{fig:ESO506-G004}, and the best-fit
parameters are summarized in Table~\ref{tab:bul_param}.

\begin{figure*}
  \epsscale{1.15}
  \plotone{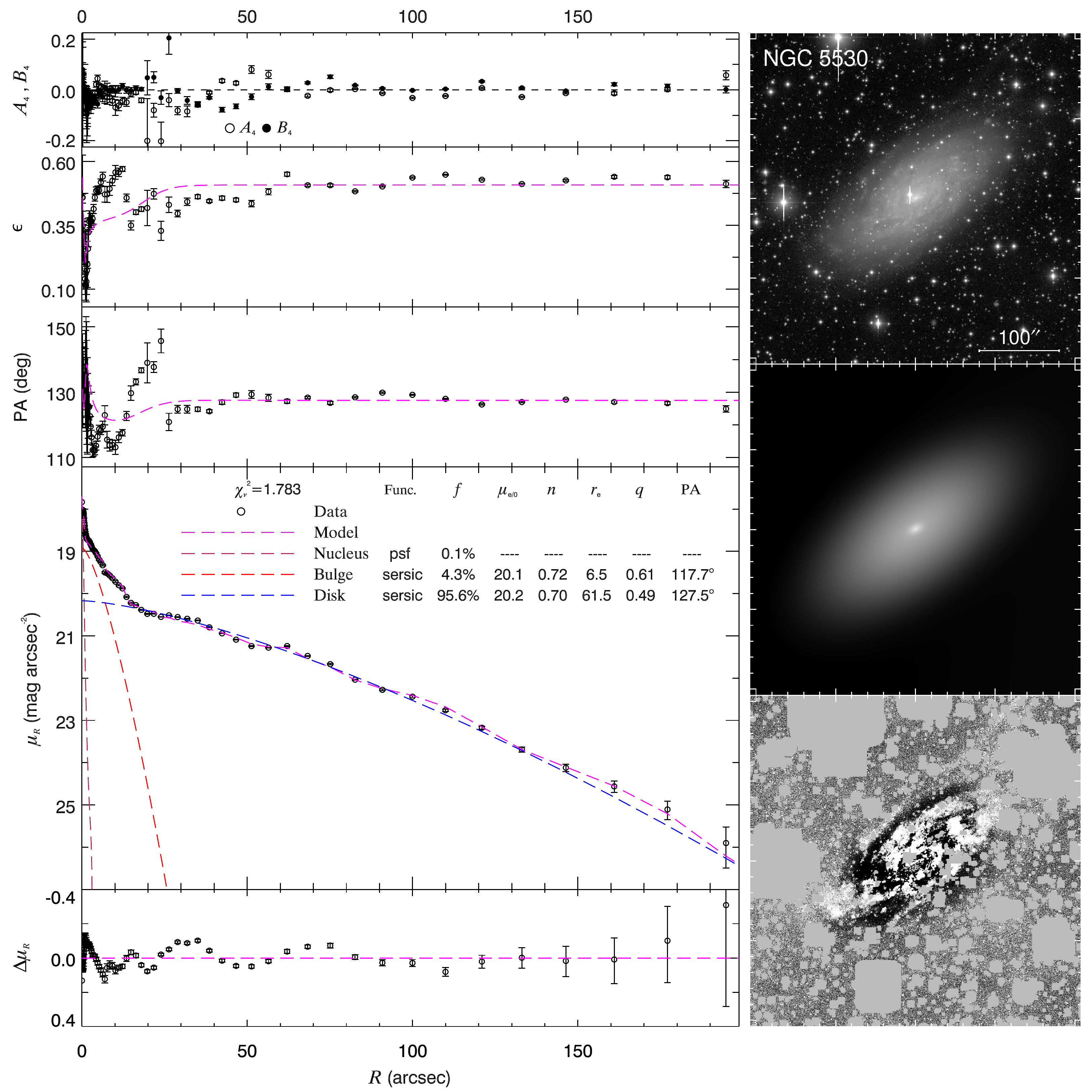}
  \caption{Best-fit model of NGC~5530, to illustrate how to model its nucleus
    and broken disk. Same convention as in
    Figure~\ref{fig:ESO506-G004}. \label{fig:NGC5530}}
\end{figure*}

\begin{figure*}
  \epsscale{1.15}
  \plotone{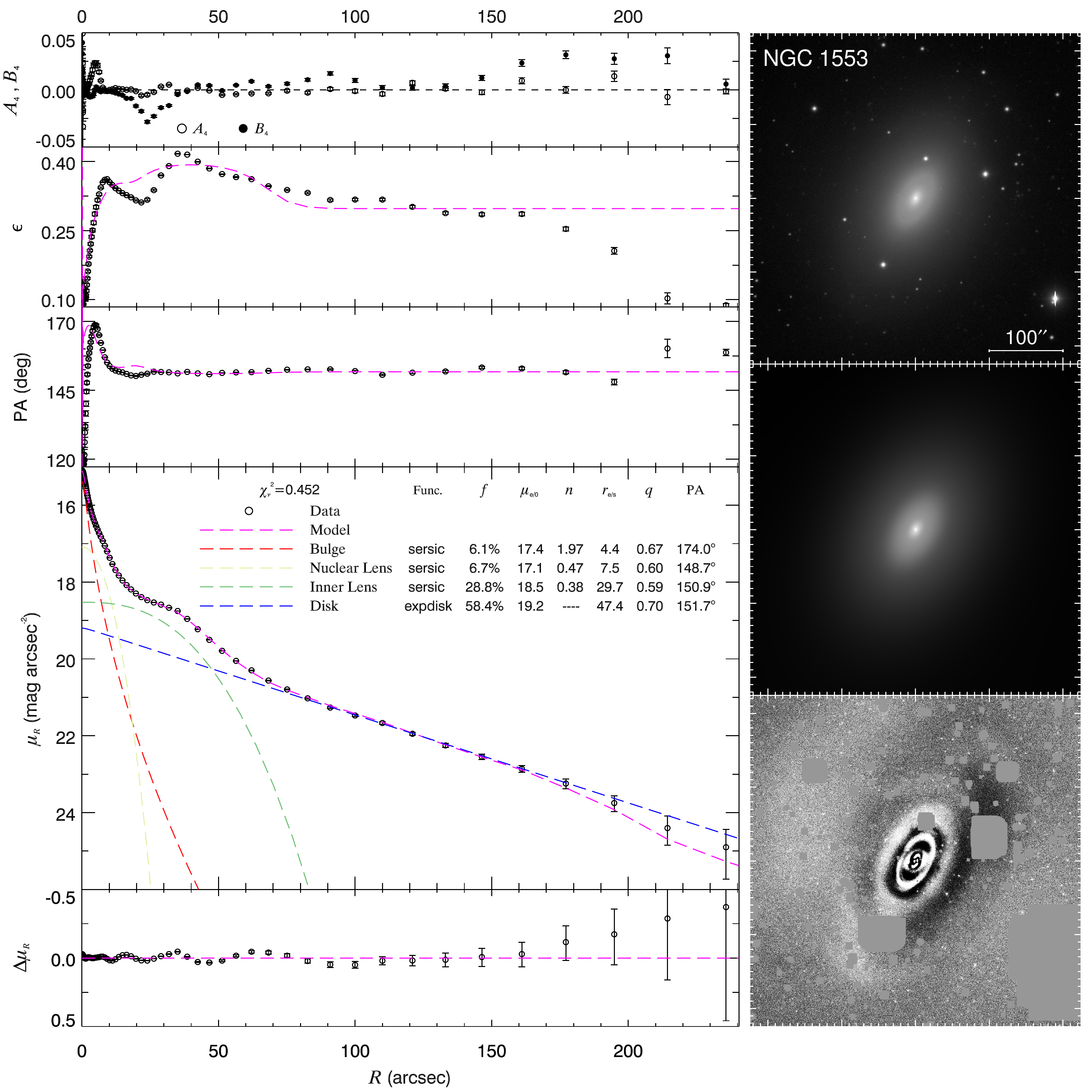}
  \caption{Best-fit model of NGC~1553, to illustrate how to model lenses. Same
    convention as in Figure~\ref{fig:ESO506-G004}. \label{fig:NGC1553}}
\end{figure*}

\begin{figure*}
  \epsscale{1.15}
  \plotone{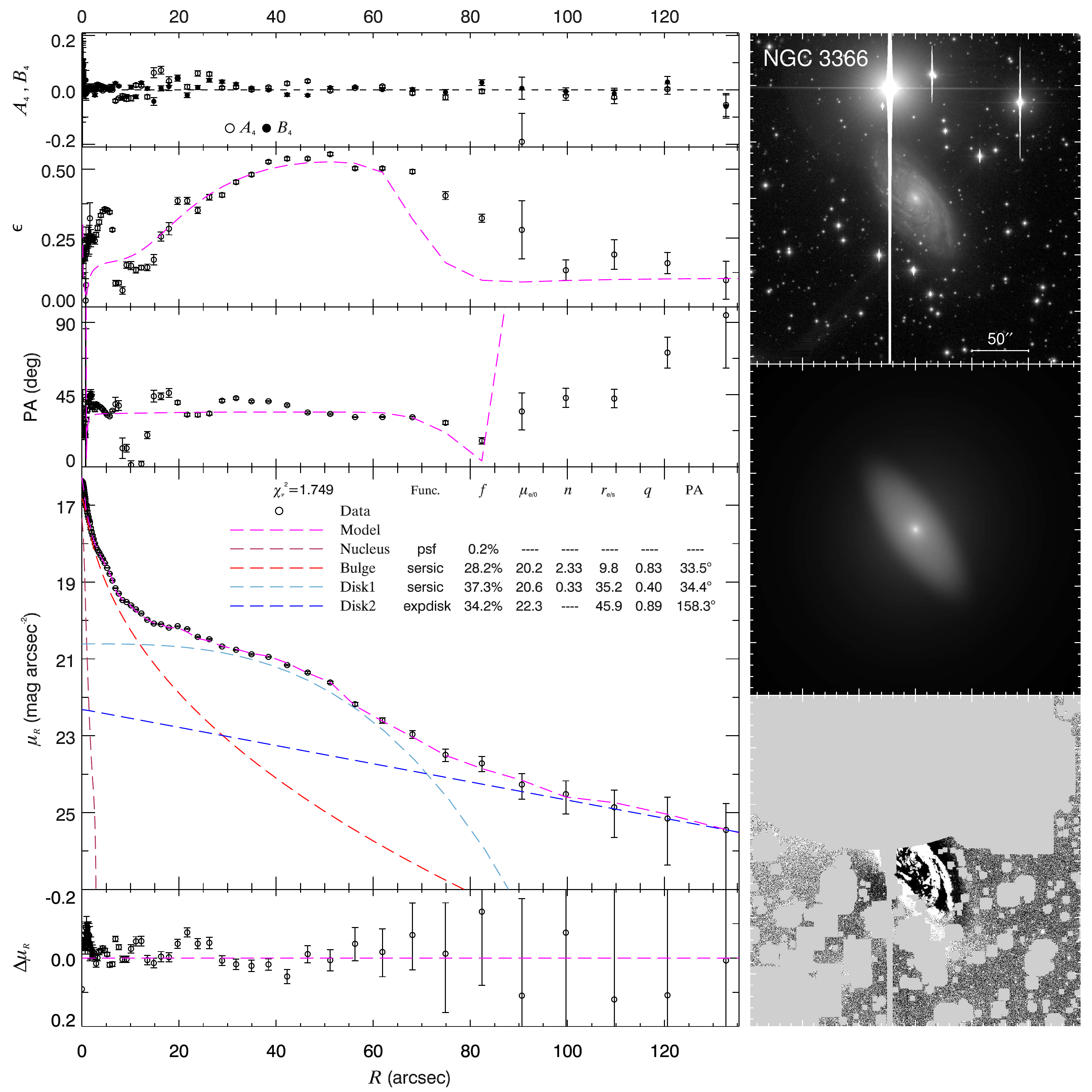}
  \caption{Best-fit model of NGC~3366, which has a two-disk configuration. Same
    convention as in Figure~\ref{fig:ESO506-G004}. \label{fig:NGC3366}}
\end{figure*}

\begin{figure*}
  \epsscale{1.15}
  \plotone{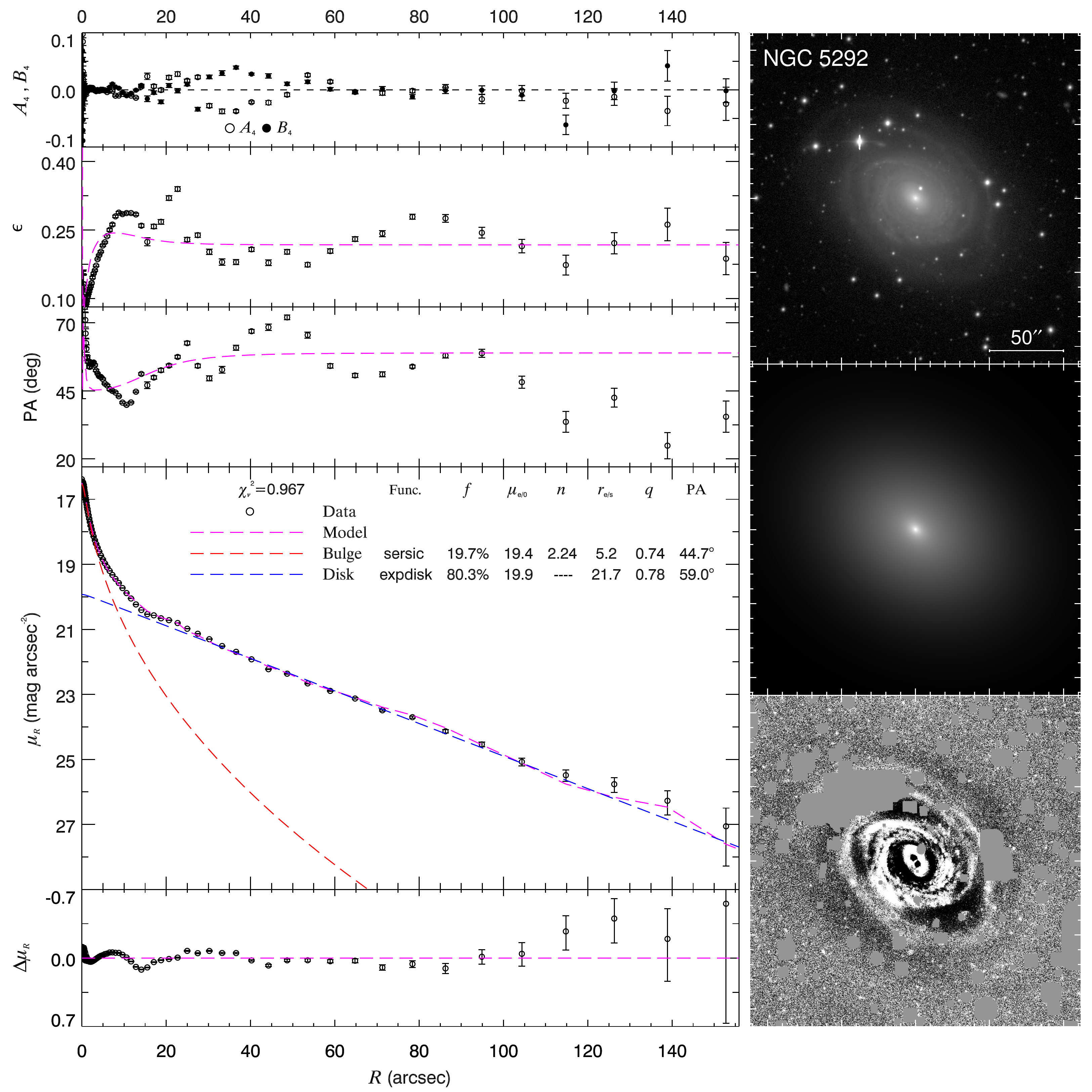}
  \caption{Best-fit model of NGC~5292, which has the simplest configuration of
    just a \sersic{} bulge and an exponential disk. Same convention as in
    Figure~\ref{fig:ESO506-G004}. \label{fig:NGC5292}}
\end{figure*}

\begin{figure*}
  \epsscale{1.15}
  \plotone{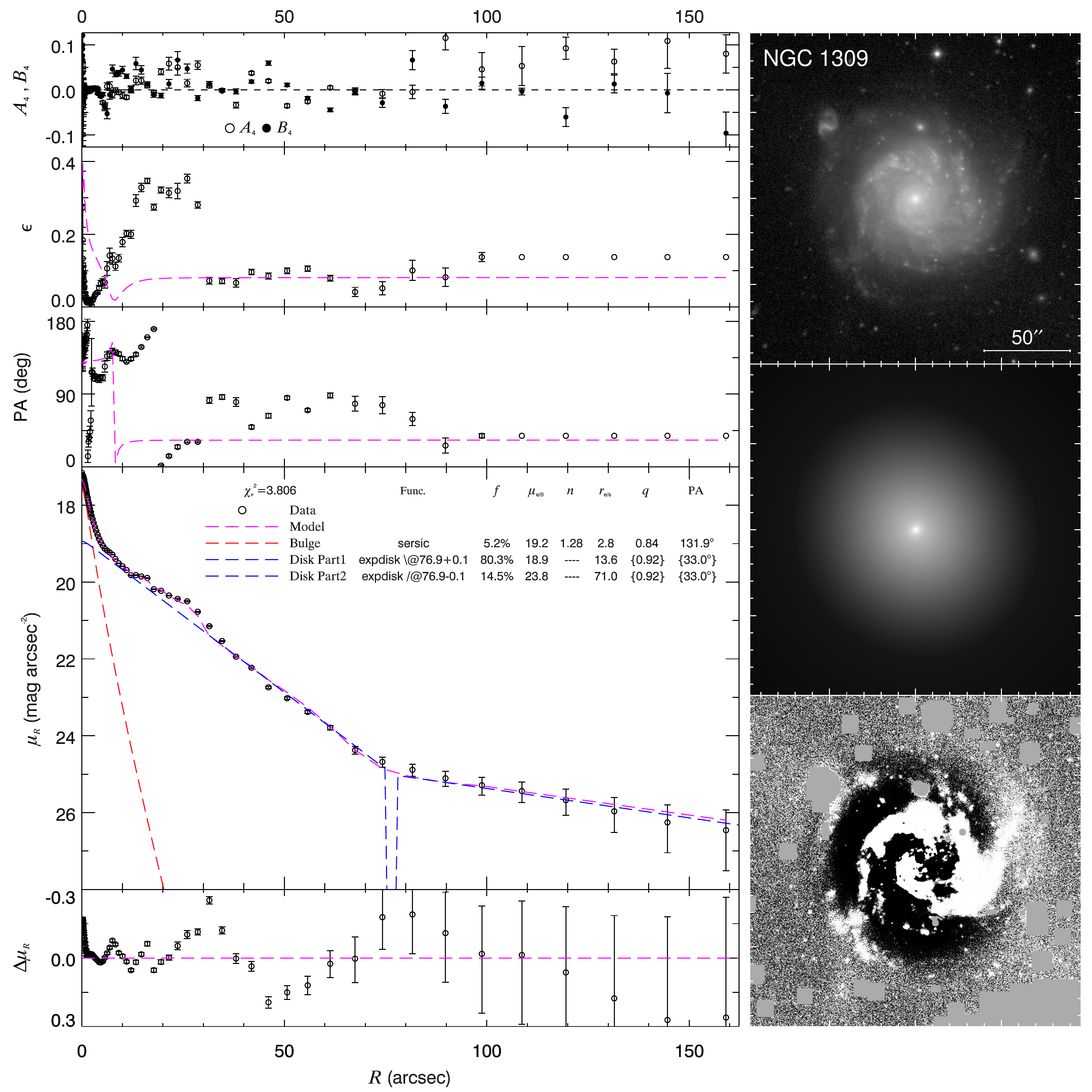}
  \caption{Best-fit model of NGC~1309, which has a Type~\Rmnum{3} disk. Same
    convention as in Figure~\ref{fig:ESO506-G004}. \label{fig:NGC1309}}
\end{figure*}

Although \galfit{} is a feature-rich tool, only part of its functionality is
needed to model the aforementioned features. Here we only provide necessary 
details of the adopted radial profiles and azimuthal functions
and refer readers to \citet{2010AJ+Peng} and Section~3.2 of \citet{2017ApJ+Gao}
for detailed descriptions and illustrations of them. The nucleus is represented
by the user-provided PSF and therefore does not have an analytic functional
form (see Figure~\ref{fig:NGC5530} for an example).  Following common practice,
we model the radial profile of the bulge using the \sersic{} function,
\begin{equation}
\label{eq:sersic}
\Sigma\left(r\right)=\Sigma_{{e}}\exp\left[-\kappa\left(\left(\frac{r}{r_{{e}}}\right)^{1/n}-1\right)\right],
\end{equation}
where \(r_{{e}}\) is the effective radius, \(\Sigma_{{e}}\) is the surface
brightness at \(r_{{e}}\), and \(n\) is the \sersic{} index; \(\kappa\) is
related to \(n\) by the incomplete gamma function,
\(\Gamma\left(2n\right)=2\gamma\left(2n,\kappa\right)\) \citep{1991A&A+Ciotti}.
We also use the \sersic{} function, usually with $n<1$, to represent lenses 
or ovals\footnote{We do not distinguish between lenses and ovals in disk 
galaxies.} (Figure~\ref{fig:NGC1553}), or disk subcomponents that have shallow 
light profiles (Figure~\ref{fig:NGC3366}).  When $n=1$, the \sersic{} function
is simply the exponential profile of a disk,
\begin{equation}
\label{eq:expdisk}
\Sigma\left(r\right)=\Sigma_{0}\exp\left(-\frac{r}{r_{{s}}}\right),
\end{equation}
where \(\Sigma_{0}\) and \(r_{{s}}\) are the central surface brightness and
scale length, respectively. Figure~\ref{fig:NGC5292} illustrates the simplest
model configuration in our study: \sersic{} bulge+exponential disk. When the
exponential profile does not describe the disk component well, for instance when
disk breaks, lenses, or rings are present, it can be substituted by a more
general \sersic{} profile (e.g., Figure~\ref{fig:NGC5530}). If a single function
does not suffice, we use a combination of profiles to represent the disk. For
instance, a combination of two truncated exponential functions with the same
orientation and ellipticity can be used to model Type~\Rmnum{2}
(Figure~\ref{fig:ESO506-G004}) and Type~\Rmnum{3} (Figure~\ref{fig:NGC1309})
disks\footnote{The Type~\Rmnum{2} (down-bending) and Type~\Rmnum{3} (up-bending)
  disks have surface brightness profiles deviating from the exponential profile
  (Type~\Rmnum{1}; \citealp[e.g.,][]{1970ApJ+Freeman,2005ApJ+Erwin,2008AJ+Erwin,
    2006A&A+Pohlen}).}. A combination of functions of different types with
possibly different orientations and ellipticities can be used to model
photometrically distinct disk subcomponents, including extra disks
(Figure~\ref{fig:NGC3366}) and lenses (Figure~\ref{fig:NGC1553}). Bars are
described by the modified Ferrer profile (e.g., Figure~\ref{fig:ESO506-G004};
\citealp{1987gady+Binney}),
\begin{equation}
\label{eq:ferrer}
\Sigma\left(r\right)=\Sigma_{0}\left[1-\left(r/r_{\mathrm{out}}\right)^{2-\beta}\right]^{\alpha}, 
\end{equation}
where \(\Sigma_{0}\) is the central surface brightness, \(\alpha\) governs the
sharpness of the outer truncation, \(\beta\) describes the central flatness of
the profile, and $r_{\mathrm{out}}$ is the radius where the profile drops to 0
and remains 0 beyond it. We generally follow \citet{2017ApJ+Gao} and let 
$\alpha$ and $\beta$ free unless they go over the allowed range 
$0\leq\alpha\leq5$. To be conservative, we fix $\alpha$ to 2 and $\beta$ to 0 
for weak bars.  We fit the sky background simultaneously with the galaxy.  The 
sky, represented by a first-order bivariate polynomial, is
\begin{equation}
\label{eq:sky}
\Sigma_{\mathrm{sky}}\left(x,y\right)=\Sigma_{\mathrm{sky}}\left(x_{{c}},y_{{c}}\right)+\left(x-x_{{c}}\right)
  \frac{\mathrm{d}\Sigma_{\mathrm{sky}}}{\mathrm{d}x}+\left(y-y_{{c}}\right)\frac{\mathrm{d}\Sigma_{\mathrm{sky}}}{\mathrm{d}y},
\end{equation}
where \(\left(x_{{c}},y_{{c}}\right)\) is the geometric center of the image and
\(\mathrm{d}\Sigma_{\mathrm{sky}}/\mathrm{d}x\) and
\(\mathrm{d}\Sigma_{\mathrm{sky}}/\mathrm{d}y\) are the sky flux gradients along
each dimension of the image. To be cautious, for galaxies that are angularly
large we fix the sky component to the sky level measured via the direct approach
described in Appendix~B.1 of \citet{2017ApJ+Gao}.

The default azimuthal shape of each galaxy component is an ellipse,
\begin{equation}
  \label{eq:pure_ell}
    r\left(x,y\right)=\left[\left(x-x_{0}\right)^{2}+\left(\frac{y-y_{0}}{q}\right)^{2}\right]^{1/2},
\end{equation}
where \(\left(x_{0},y_{0}\right)\) is the centroid of the ellipse, the $x$-axis
is aligned with the major axis of the ellipse, and \(q\) is the axis ratio. The
Fourier modes perturb the ellipse in a way described by
\begin{equation}
  \label{eq:fourier_mode}
  r\left(x,y\right)=r_{0}\left(x,y\right)\left(1+\sum_{m=1}^{N}a_{m}\cos\left[m\left(\theta+\phi_{m}\right)\right]\right),
\end{equation}
where \(r_{0}\) is the unperturbed radius, \(a_{m}\) is the amplitude for mode
\(m\), \(\theta=\arctan\left((y-y_{0})/\left((x-x_{0})q\right)\right)\), and
\(\phi_{m}\) is the phase angle relative to \(\theta\). The Fourier modes are
generally not invoked, except for cases that need to model boxy/peanut bulges
($m=4$; Figure~\ref{fig:ESO506-G004}) or lopsided disks ($m=1$). Alternatively,
the bending modes can induce curvature by only perturbing the $y$-axis following
\begin{equation}
  \label{eq:bending_mode}
  y^{\prime}=y+\sum_{m=1}^{N}a_{m}\left(\frac{x}{r_{\mathrm{scale}}}\right)^{m},
\end{equation}
where $a_{m}$ is similar to the one in Fourier modes and $r_{\mathrm{scale}}$ is
the scale radius of the corresponding radial profile (e.g., $r_{e}$ for
\sersic{} function and $r_{s}$ for exponential function). Note that we apply the
bending modes only once in this study, in order to model the twisted bar of
IC~4618 ($m=2$).

The truncation function can alter both the radial profile and azimuthal shape of
components. We restrict its applications to create composite profiles to model
disk breaks as well as rings (see Figures~\ref{fig:ESO506-G004} and
\ref{fig:NGC1309} for examples). Such a composite profile has an inner part
described by a certain analytic function and an outer part that behaves as
another, and they are modified by the same truncation function, albeit in
opposite manners (outer truncation and inner truncation). In such cases, the
truncation function does not carry any physical meaning but only serves to link
the inner and outer parts. Moreover, the overlap region of the two parts can
naturally produce ring-like features. The truncation function is basically a
hyperbolic tangent function, and its functional dependence on various parameters
is given schematically by
\begin{equation}
\label{eq:truncation}
P\left(x,y\right)=\tanh\left(x,y;x_{0},y_{0},r_{\mathrm{break}},\bigtriangleup r_{\mathrm{soft}},q,\theta_{\mathrm{PA}}\right),
\end{equation}
where \((x_{0},y_{0})\) is the center, \(q\) is the axis ratio, and
\(\theta_{\mathrm{PA}}\) is the position angle of the truncation function.
These three parameters are hidden by default; if not specified, their values are
inherited from the component that is modified by the truncation function.  The
break radius \(r_{\mathrm{break}}\) marks the location where the flux of the
truncated model drops to 99\% of its original flux.  The softening length
\(\bigtriangleup r_{\mathrm{soft}}\) is defined as
\(r_{\mathrm{soft}}-r_{\mathrm{break}}\) or
\(r_{\mathrm{break}}-r_{\mathrm{soft}}\) for outer truncation or inner
truncation, respectively, where \(r_{\mathrm{soft}}\) is the radius where the
truncated model flux drops to 1\% of its original flux. Its detailed analytic
form is lengthy and is not of immediate interest; readers can consult Appendix~B
in \citet{2010AJ+Peng} for details. Components are modified by the truncation
function by multiplying the original flux distribution with \(P\) for inner
truncation and with \(1-P\) for outer truncation.

\input{./figset}

\begin{deluxetable*}{lCCCCCCCCc}
  \tabletypesize{\scriptsize}
  \tablecaption{Best-fit Parameters and Bar/Lens Identifications of the Disk
    Galaxies \label{tab:bul_param}}

  \tablehead{\colhead{Name} & \colhead{$m$} & \colhead{$B/T$} &
    \colhead{$\mu_{e}$} & \colhead{$n$} & \colhead{$r_{e}$} &
    \colhead{$\epsilon$} & \colhead{$D/T$} & \colhead{$b/T$} &
    \colhead{Bar/Lens} \\ \colhead{} & \colhead{(mag)} & \colhead{} &
    \colhead{(mag~arcsec$^{-2}$)} & \colhead{} & \colhead{(arcsec)} & \colhead{}
    & \colhead{} & \colhead{} &  \colhead{} \\
    \colhead{(1)} & \colhead{(2)} & \colhead{(3)} & \colhead{(4)} &
    \colhead{(5)} & \colhead{(6)} & \colhead{(7)} & \colhead{(8)} &
    \colhead{(9)} & \colhead{(10)}}

  \startdata
  ESO 027--G001 & 14.81\pm 0.11 & 0.045\pm 0.005 & 19.03\pm 0.18 & 3.44\pm 0.34 & \phn 2.30\pm  0.20 & 0.564\pm 0.051 & 0.890 & 0.065 & B \\
  ESO 121--G026 & 14.00\pm 0.11 & 0.091\pm 0.008 & 18.68\pm 0.16 & 1.58\pm 0.13 & \phn 2.54\pm  0.20 & 0.214\pm 0.018 & 0.825 & 0.084 & B \\
  ESO 137--G034 & 12.97\pm 0.09 & 0.263\pm 0.021 & 19.79\pm 0.15 & 1.97\pm 0.16 & \phn 6.61\pm  0.46 & 0.245\pm 0.019 & 0.737 & 0.000 & N \\
  ESO 138--G010 & 13.56\pm 0.35 & 0.066\pm 0.017 & 21.64\pm 0.63 & 2.14\pm 0.68 & 11.60\pm  4.46 & 0.252\pm 0.005 & 0.930 & 0.000 & N \\
  ESO 186--G062 & 14.07\pm 0.23 & 0.139\pm 0.022 & 22.66\pm 0.51 & 3.28\pm 0.57 & 19.74\pm  5.79 & 0.660\pm 0.051 & 0.840 & 0.021 & B \\
  ESO 213--G011 & 13.90\pm 0.17 & 0.085\pm 0.012 & 21.34\pm 0.28 & 2.36\pm 0.30 & \phn 8.71\pm  1.34 & 0.293\pm 0.023 & 0.915 & 0.000 & N \\
  ESO 221--G026 & 11.13\pm 0.17 & 0.534\pm 0.062 & 19.46\pm 0.46 & 5.00\pm 0.65 & 13.48\pm  3.72 & 0.528\pm 0.041 & 0.466 & 0.000 & ? \\
  ESO 221--G032 & 15.50\pm 0.17 & 0.029\pm 0.004 & 18.44\pm 0.28 & 2.35\pm 0.30 & \phn 1.03\pm  0.16 & 0.202\pm 0.016 & 0.971 & 0.000 & N \\
  ESO 269--G057 & 12.79\pm 0.10 & 0.207\pm 0.024 & 18.10\pm 0.17 & 1.37\pm 0.14 & \phn 3.68\pm  0.32 & 0.287\pm 0.026 & 0.724 & 0.069 & B \\
  ESO 271--G010 & 16.41\pm 0.10 & 0.014\pm 0.001 & 21.54\pm 0.15 & 1.48\pm 0.12 & \phn 4.36\pm  0.31 & 0.586\pm 0.045 & 0.948 & 0.038 & W \\
  ESO 320--G026 & 14.51\pm 0.17 & 0.070\pm 0.009 & 17.96\pm 0.28 & 0.98\pm 0.13 & \phn 1.97\pm  0.30 & 0.478\pm 0.037 & 0.930 & 0.000 & N \\
  ESO 321--G025 & 16.83\pm 0.10 & 0.012\pm 0.001 & 19.76\pm 0.15 & 0.57\pm 0.05 & \phn 2.37\pm  0.17 & 0.720\pm 0.056 & 0.954 & 0.034 & W \\
  ESO 380--G001 & 13.98\pm 0.11 & 0.081\pm 0.008 & 19.57\pm 0.18 & 3.94\pm 0.39 & \phn 3.71\pm  0.33 & 0.446\pm 0.041 & 0.741 & 0.178 & B \\
  ESO 380--G006 & 12.23\pm 0.13 & 0.284\pm 0.032 & 20.52\pm 0.21 & 2.86\pm 0.30 & 12.33\pm  1.48 & 0.298\pm 0.027 & 0.716 & 0.000 & L \\
  \enddata

  \tablecomments{Col.~(1): Galaxy name. Cols.~(2)--(7): Total $R$-band magnitude
    of the bulge, bulge-to-total flux ratio, bulge surface brightness at the
    effective radius, bulge \sersic{} index, bulge effective radius, and bulge
    ellipticity.  Col.~(8): Disk-to-total ratio. Col.~(9): Bar-to-total ratio.
    Col.~(10): Flag for the presence or absence of a bar/lens: B = definitely
    barred; W = weakly barred; N = no bar or lens; L = no bar but lens present;
    ? = uncertain. \\ (Table~\ref{tab:bul_param} is published in its entirety in
    machine-readable format. A portion is shown here for guidance regarding its
    form and content.)}
\end{deluxetable*}

\subsection{Uncertainties of Bulge Parameters}
\label{sec:uncert-best-fit}

As CGS galaxies are bright and well-resolved, the noise and PSF contribute 
negligibly to the uncertainties of the bulge parameters (total magnitude $m$, 
bulge-to-total ratio $B/T$, effective surface brightness $\mu_{e}$, \sersic{} 
index $n$, effective radius $r_{e}$, and ellipticity $\epsilon$).  The major 
source of uncertainty comes from sky level measurements and, more importantly,
model assumptions. We measure the sky-induced uncertainties as variations of the
best-fit bulge parameters when perturbing the sky levels around 
$\pm1\sigma_{\mathrm{sky}}$ of the best-fit sky levels, where uncertainties of 
the sky level $\sigma_{\mathrm{sky}}$ are measured as the root-mean-square of 
the residuals measured from randomly placed boxes in the sky-dominated region 
of the sky-subtracted data image (see Appendix~B of \citealp{2017ApJ+Gao} for 
details).

One source of systematic uncertainty comes from omission of certain features of
the galaxy in the model. The effects were estimated by \citet{2017ApJ+Gao} by 
comparing bulge parameters from input models with and without the features. 
We repeat their conclusions here. Ignoring outer lenses/rings will induce 
uncertainties of 0.05\,mag, 7.1\%, 0.09\,mag~arcsec$^{-2}$, 5.8\%, 5.3\%, and 
4.8\% for $m$, $B/T$, $\mu_{e}$, $n$, $r_{e}$, and $\epsilon$, respectively. 
Spiral arms in barred and unbarred galaxies impact bulge parameters in different
manners, since spiral arms usually stop at the ends of the bar in barred 
galaxies, whereas they extend to the center in unbarred systems.  Not 
including spiral arms in the model introduces uncertainties at the level of
0.14\,mag, 11.7\%, 0.24\,mag~arcsec$^{-2}$, 10.1\%, 13.6\%, and 0.4\% for $m$,
$B/T$, $\mu_{e}$, $n$, $r_{e}$, and $\epsilon$, respectively, for the bulges 
of unbarred galaxies; for barred galaxies, the corresponding values are 
0.03\,mag, 2.2\%, 0.03\,mag~arcsec$^{-2}$, 1.1\%, 1.4\% and 0.4\% for $m$, 
$B/T$, $\mu_{e}$, $n$, $r_{e}$, and $\epsilon$, respectively. Another source of
systematic uncertainty stems from the use of different mathematical 
representations of the same disk surface brightness, which arises when we model
disk breaks, lenses, and rings, along with the underlying disk. The typical 
contribution to the error budget is 0.09\,mag, 6.7\%, 0.15\,mag~arcsec$^{-2}$, 
8.0\%, 6.9\%, and 7.7\% for $m$, $B/T$, $\mu_{e}$, $n$, $r_{e}$, and 
$\epsilon$, respectively. The final uncertainties of the bulge parameters in 
Table~\ref{tab:bul_param} represent the quadrature sum of the uncertainties 
from these various considerations.

\section{Comparison with S$^{4}$G}
\label{sec:comparison-with-s4g}

Some of the CGS galaxies were also observed as part of S$^{4}$G.
\citet{2015ApJS+Salo} applied \galfit{} to perform human-supervised,
multi-component decomposition of the $3.6\,\micron$ images.  In addition to
bulges and disks, they fit bars and nuclear point sources, with up to four
components in the model.

\begin{figure*}
\epsscale{1.15}
\plotone{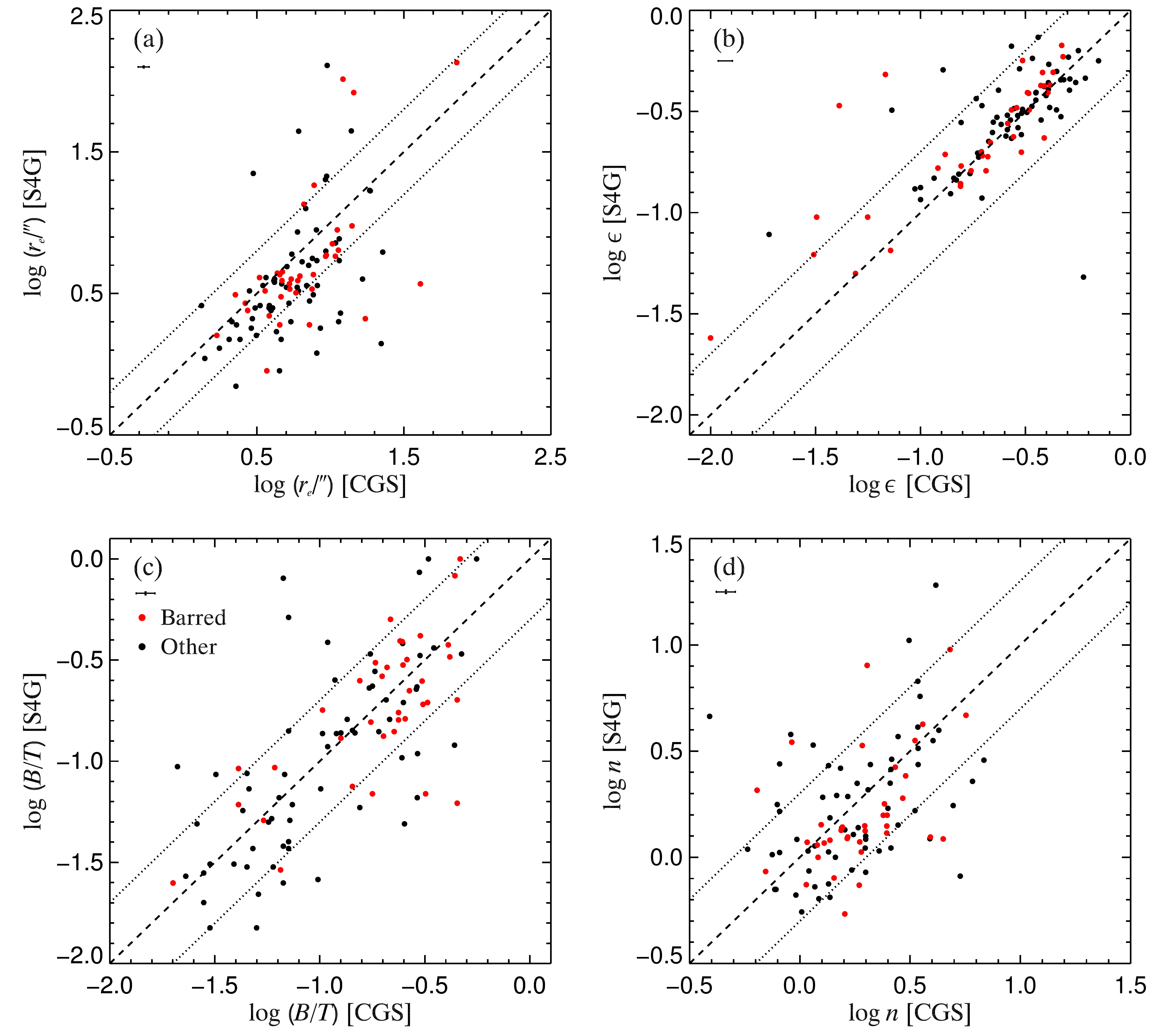}
\caption{Comparison of CGS bulge parameters with S$^{4}$G Pipeline4 results, for
  (a) effective radii, (b) apparent ellipticities, (c) bulge-to-total ratios,
  and (d) \sersic{} indices. Barred galaxies are highlighted in red. Median
  errors of the bulge parameters are illustrated in the upper-left corner of
  each panel.  The constant errors of S$^{4}$G bulge parameters are median
  errors introduced by PSF and sky subtraction, as listed in Tables~3~and~4 of
  \citet{2015ApJS+Salo}.  In each panel, the dashed line gives the one-to-one
  relation, and the dotted lines demarcate an offset of 0.3\,dex (a factor of
  2). \label{fig:comp_s4g}}
\end{figure*}

Although their philosophy in construction of multi-component models is similar
to ours, the details differ greatly.  For instance, they do not treat disk
breaks, and we do not limit our fits to a pre-determined number of components,
if more sophisticated models are deemed necessary for complex situations. It is
of interest to know how our different approaches affect the final results.  We
cross match our sample with the S$^{4}$G sample and find 101 galaxies in common
that have bulge decompositions of relatively high quality (at least 4 according
to the rating system of \citealp{2015ApJS+Salo}; see their Section~5.1).
Figure~\ref{fig:comp_s4g} compares the effective radii, apparent ellipticities,
bulge-to-total ratios, and \sersic{} indices from the two independent sets of
decompositions. We find broad consistency between our results and those of the
S$^{4}$G Pipeline4: 74\%, 74\%, 68\%, and 90\% of the bulge parameters $B/T$,
$n$, $r_{e}$, and $\epsilon$, respectively, agree with each other within a
factor of 2. But we also find many cases of S$^{4}$G-derived bulge parameters
(especially $B/T$, $r_{e}$, and $n$) that are systematically lower than our
values for barred galaxies. We suspect that this is due to the fact that disk
breaks are prevalent in barred galaxies, and failure to model the disk breaks
will underestimate the contribution from the bulge and bar
\citep[e.g.,][]{2017ApJ+Gao}.

However, some of the extreme (larger than a factor of 2) outliers are
disconcerting and require closer scrutiny.  Upon careful examination, we find
that almost all the extreme outliers can be attributed to systematic differences
in model construction.  Among the 15 extreme outliers with systematically lower
values of $B/T$ from S$^{4}$G, 11 have disk breaks that were not
treated\footnote{They are IC~1993, IC~2051, NGC~1232, NGC~1292, NGC~1452,
  NGC~1640, NGC~3673, NGC~3887, NGC~4462, NGC~7513, and NGC~7590.}. The 11
extreme outliers with S$^{4}$G $B/T$ values larger than ours can be traced to
various reasons, the most common being the presence of a nucleus (not modeled by
S$^{4}$G), which leads to a systematic overestimation of the bulge \sersic{}
index (and bulge luminosity). Other reasons include the misidentification of
disk galaxies as ellipticals and the neglect of extra disk components, such as
thick disks and lenses.  For instance, the three sources for which S$^{4}$G
derived $B/T=1$ are, in fact, misclassified as ellipticals
\citep{2013ApJ+Huang1}, and we decomposed them as S0s or spirals
\footnote{IC~2006 and NGC~3904 are S0s with lenses, and NGC~7213 is a spiral
  galaxy that shows evident spiral and ring features in the central
  30\arcsec.}. Among the 22 outliers with significantly smaller $r_{e}$ derived
from S$^{4}$G, 15 are due to disk breaks that were not taken into
account\footnote{They are IC~1953, IC~1993, IC~2051, NGC~1022, NGC~1084,
  NGC~1232, NGC~1292, NGC~1452, NGC~3673, NGC~3887, NGC~4462, NGC~4899,
  NGC~5339, NGC~7513, and NGC~7590.}. At the same time, of the 10 sources with
exceptionally large values of $r_{e}$ from S$^{4}$G, four are due to large-scale
essential components that were missing in the model\footnote{They are NGC~584,
  NGC~1302, NGC~4050, and NGC~5078.} (e.g., bars, lenses, thick disks), three
are actually S0s misclassified as ellipticals, as mentioned above, and two stem
from neglecting a nucleus\footnote{NGC~3892 and NGC~4802}. Among the 14 extreme
outliers with lower S$^{4}$G values of $n$, nine are due to disk
breaks\footnote{They are IC~1993, IC~2051, NGC~1300, NGC~1640, NGC~3673,
  NGC~4462, NGC~7140, NGC~7590, and NGC~7755.}, and two are due to modeling
lenses with an exponential instead of a low-$n$ \sersic{}
function\footnote{NGC~1425 and NGC~3885}; of the 12 objects that are positive
outliers in $n$, six are due to nuclei\footnote{They are NGC~150, NGC~4684,
  NGC~4802, NGC~4965, NGC~5339, and NGC~7531.}, and four stem from large-scale
essential components that are missing in model\footnote{They are NGC~4050,
  NGC~5078, NGC~5468, and NGC~7213.}  (e.g., bars, disks, and disk
subcomponents). In terms of $\epsilon$, almost all of the extreme outliers are
biased too high in S$^{4}$G.  One-third of these cases are caused by missing
components with high ellipticities near the bulge, such as bars, bar-like
patterns produced by winding spiral arms, and additional disk components. Some
are due to the effect of nuclei, which, when neglected, leads to larger $n$ and
$r_{e}$ for the bulge and thus absorbs some of disk/bar light. Mistaking disks
for ellipticals also results in larger $\epsilon$ (e.g., NGC~7213). The only
single extreme outlier (NGC~4684) with a very low value of $\epsilon$ in
S$^{4}$G is due to modeling its nucleus as the bulge.

To summarize: the above comparison re-emphasizes our motivation to provide a new
set of revised bulge parameters for nearby galaxies. Even though S$^{4}$G has
already performed very careful, highly sophisticated multi-component
decomposition, detailed comparison with our analysis reveals that significant
discrepancies can still arise.  The uncertainties of the bulge parameters are
dominated entirely by systematic differences in model construction; wavelength
effects play a minor role. As \citet{2017ApJ+Gao} stress, proper treatment of
certain secondary morphological components are absolutely
indispensable---indeed, obligatory---if one wishes to obtain robust structural
parameters for the bulge. The external comparison here further strengthens their
conclusions.

\section{Bulge Prominence along the Hubble Sequence}
\label{sec:bulge-prom-along}

Bulge prominence is one of the key defining criteria of the Hubble
classification scheme \citep{1926ApJ+Hubble,1936rene+Hubble,1961hag+Sandage}.
Therefore, it is expected that bulge properties should correlate with Hubble
types to some extent. Huge effort has been devoted to investigate whether Hubble
types are good predictors of bulge prominence and vice versa
\citep[e.g.,][]{1985ApJS+Kent,1986ApJS+Kodaira,1986ApJ+Simien,1989AJ+Solanes}.
Many studies have shown that $B/T$ does correlate with morphological type index
$T$ in an average sense, with minor counterarguments \citep{1986ApJS+Kodaira,
  1992PhDT+Byun,2004A&A+Grosbol}. However, previous studies were either limited
by small sample size \citep[e.g.,][]{1995ApL&C+Heraudeau,1996A&A+de_Jong3,
  2000ApJ+Khosroshahi,2001AJ+Graham,2001A&A+Mollenhoff,2004A&A+Mollenhoff},
over-simplified decomposition techniques \citep[e.g.,][]{1985ApJS+Kent,
  1986ApJ+Simien,1986ApJS+Kodaira,2001AJ+Graham}, or incorrect assumptions of
the bulge profile \citep[e.g.,][]{1985ApJS+Kent,1986ApJ+Simien,1986ApJS+Kodaira,
  2009ApJ+Oohama}. Subsequent efforts addressed some of these shortcomings
\citep{1996A&A+de_Jong3,2007MNRAS+Laurikainen,2010MNRAS+Laurikainen,
  2009ApJ+Weinzirl,2017A&A+Mendez-Abreu}. This study represents a significant
contribution toward these efforts, with the employment of well-developed 2D
techniques, more realistic model assumptions, and better understanding of the
error budget. Using the vastly improved bulge measurements for a sizable sample
presented in this paper, we revisit this classical problem and showcase the
potential of our database.

\begin{figure}
  \epsscale{1.17}
  \plotone{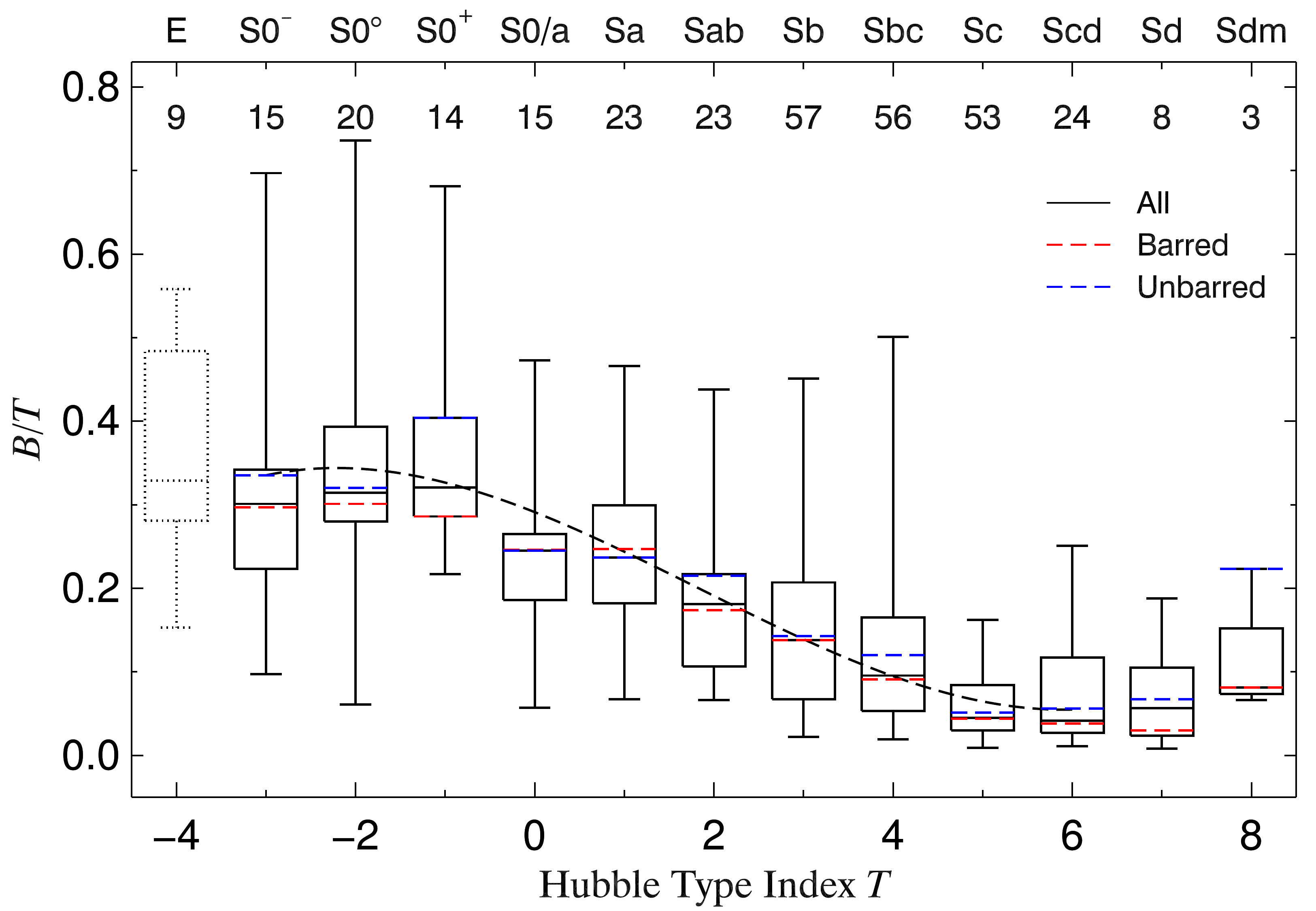}
  \caption{Distribution of $B/T$ as a function of Hubble type. The morphological
    type index $T$ is given in the bottom axis and the corresponding traditional
    Hubble types are given in the top axis. For each bin in morphological type,
    the distribution of $B/T$ is described by a box-and-whisker plot, for which
    the box encloses the interquartile range and the whisker indicates its
    maximum and minimum. Note that we group all the S0s that were misclassified
    as ellipticals into the $T=-4$ bin (see Section~\ref{sec:sample-data}) and
    distinguish them with the dotted box and whisker. The horizontal lines of
    each morphological bin represent the median for all galaxies (black), barred
    galaxies (red dashed), and unbarred galaxies (blue dashed). The number of
    galaxies in each bin is given on top of the whisker. The dashed line
    represents the polynomial fit (Equation~\ref{eq:frac_T}) to the mean $B/T$
    in Hubble type bins with more than 10 galaxies. \label{fig:bul_T}}
\end{figure}

Figure~\ref{fig:bul_T} shows the distribution of $B/T$ in morphological bins
from S0 to Sdm. We confirm previous findings that median $B/T$ decreases toward
late Hubble types. For individual galaxies, the Spearman's rank correlation
coefficient between $B/T$ and morphological type index $T$ is $-0.71$, and the
correlation is statistically significant ($p$-value is 0.00). Excluding bins
with fewer than 10 galaxies, the best-fit third-order polynomial to the mean
$B/T$ as a function of $T$ is
\begin{eqnarray}
  \label{eq:frac_T}
  \langle B/T \rangle&=& 0.29\pm0.01-(0.042\pm0.006)T-(0.006\pm0.003)T^2 \nonumber\\
  & & +(0.0011\pm0.0004)T^3.
\end{eqnarray}
The three bins of
S0 galaxies have roughly constant $B/T$ \citep[see also][]{1986ApJ+Simien,
  2000ApJ+Khosroshahi,2007MNRAS+Laurikainen,2010MNRAS+Laurikainen}. Although the
overall trend is similar to that of previous studies, it is worthwhile to note
that many authors find that $B/T$ systematically decreases for Hubble types
later than $T=4$ \citep[e.g.,][]{1986ApJ+Simien,1995ApL&C+Heraudeau,
  2008MNRAS+Graham,2017A&A+Mendez-Abreu}, but we do not\footnote{For similar
  results, see \citealp{2001AJ+Graham,2004A&A+Grosbol,2007MNRAS+Laurikainen,
    2010MNRAS+Laurikainen,2009ApJ+Weinzirl}.}. We do not know whether this is
genuine or a bias due to the selection of CGS galaxies against fainter galaxies
in these late Hubble type bins. We confirm that most of the bins with meaningful
statistics exhibit a large scatter in $B/T$ \citep[e.g.,][]{1985ApJS+Kent,
  1986ApJS+Kodaira,1986ApJ+Simien,2009ApJ+Weinzirl,2010MNRAS+Laurikainen}. The
scatter is especially remarkable among S0 galaxies, whose $B/T$ can be as large
as 0.7 and as small as those of Sc galaxies ($B/T \approx 0.1$; see also
\citealp{2018ApJ+Gao}). Other classification criteria in the Hubble sequence,
such as properties of spiral arms, and classification errors may be responsible
for the scatter present in all Hubble types. The large dispersion of bulge
prominence at any given Hubble type precludes the use of Hubble type to
quantitatively predict $B/T$.

Despite the general agreement of the systematic trend and scatter of $B/T$ along
the Hubble sequence, we note that studies that employ 1D techniques and
classical models (i.e., a de~Vaucouleurs bulge and an exponential disk)
systematically overestimate $B/T$ compared with our 2D multicomponent
decomposition. For example, \citet{1986ApJ+Simien}, \citet{1985ApJS+Kent}, and
\citet{1986ApJS+Kodaira} measured optical $\langle B/T\rangle \ga 0.5$ of S0s;
\citet{1986ApJ+Simien} and \citet{1985ApJS+Kent} found no S0 with $B/T$ smaller
than $\sim0.3$. By contrast, we obtain $\langle B/T\rangle=0.34\pm0.15$ in
$R$-band.  Less dramatic, though still significant, overestimates of bulge flux
are also seen in later Hubble types. The de~Vaucouleurs law ($n=4$) has long
been proven to be inadequate for most disk galaxies (see Figure~2b of
\citealp{2018ApJ+Gao}), especially those of late-type
\citep{1994MNRAS+Andredakis,1995MNRAS+Andredakis,1996A&A+de_Jong2}. Application
of a universal de~Vaucouleurs law to extract the bulges of all galaxy types will
lead to systematic overestimates of the bulge flux
\citep[e.g.,][]{2009ApJ+Oohama}.

Finally, we divide the sample into barred and unbarred galaxies to examine their
potential difference in bulge properties. Apart from S0s (see Section~4.1 of
\citealp{2018ApJ+Gao}), we find that the barred and unbarred galaxies have
similar median $B/T$ along the Hubble sequence, in contrast with some previous
studies that show systematically weaker bulges in barred galaxies
\citep[e.g,][]{2007MNRAS+Laurikainen,2009ApJ+Weinzirl}. We attribute the
discrepancy to our more accurate decomposition that consistently accounts for
disk breaks in barred galaxies, which, if neglected, leads to underestimation of
the bulge flux.

\section{Summary}
\label{sec:summary}

We perform 2D multi-component decompositions of 258 CGS spiral galaxies in $R$
band. In addition to bulges and disks, we successfully model nuclei, bars, disk
breaks, nuclear/inner lenses, and inner rings.  Our decomposition intentionally
ignores nuclear rings/bars, which we consider to be part of the photometric
bulge, and we also do not treat spiral arms, outer lenses and outer rings
because they are known to be unimportant for bulge measurements. We pay close
attention to estimating robust errors for the derived bulge parameters, taking
into account the uncertainties from sky level measurements and model
assumptions. Together with the 62 CGS S0s separately analyzed by
\citet{2018ApJ+Gao}, we present a homogeneous catalog of bulge parameters for
320 CGS disk galaxies. Comparison of our bulge parameters with the results from
S$^{4}$G shows significant discrepancies that cannot be accounted for by
wavelength effects. We find that differences in model assumptions is the major
source of the inconsistency, stressing the need to construct realistic models
that consider all the necessary secondary morphological features in the image
decomposition \citep{2017ApJ+Gao}.

We reevaluate the classic relation between bulge prominence and Hubble type,
confirming that, while $B/T$ decreases systematically from early to late-type
disk galaxies, the scatter in $B/T$ is considerable at any given morphological
type. In contrast with previous studies that claim barred galaxies host weaker
bulges, we show that barred and unbarred galaxies have similar median $B/T$
across the Hubble sequence except for S0s.

The catalog of bulge parameters presented here is a homogeneous and robust
dataset, one that has the promise for new discoveries.  Detailed analysis of the
products, including statistics and correlation of bulge parameters, their
scaling relations, and study of bulge types, will be presented in forthcoming
papers.

\acknowledgments

This work was supported by the National Science Foundation of China (11721303)
and the National Key R\&D Program of China (2016YFA0400702). ZYL is supported by
the Youth Innovation Promotion Association, Chinese Academy of Sciences. This
research has made use of the NASA/IPAC Extragalactic Database (NED) which is
operated by the Jet Propulsion Laboratory, California Institute of Technology,
under contract with the National Aeronautics and Space Administration.

\appendix

\section{The Bulgeless Galaxies}
\label{sec:bulgeless-galaxies}

We tabulate the bulgeless galaxies and the presence of nuclei therein in
Table~\ref{tab:bul_less}.

\begin{deluxetable}{lcccc}
  \tabletypesize{\footnotesize}
  \tablecaption{Basic Properties of the Bulgeless Galaxies \label{tab:bul_less}}

  \tablehead{\colhead{Name} & \colhead{$T$} & \colhead{$\log \,
      (M_{\star}/M_{\sun})$} & \colhead{Bar} & \colhead{Nucleus} \\
    \colhead{(1)} & \colhead{(2)} & \colhead{(3)} & \colhead{(4)} &
    \colhead{(5)}}

  \decimals
  \startdata
  ESO 383--G087 & 7.9 & \phn 7.78 &  N  & \nodata \\      
  ESO 445--G089 & 6.7 & \phn 9.60 &  B  & \nodata \\
  IC 4710      & 8.9 & \phn 8.48 &  N  & NSC\tablenotemark{a} \\
  IC 5201      & 6.1 & \phn 8.76 &  B  & \ion{H}{2}\tablenotemark{b} \\
  NGC 45       & 7.8 & \phn 9.07 &  N  & X-ray\tablenotemark{c} \\
  NGC 247      & 6.9 & \phn 8.97 &  N  & NSC\tablenotemark{a} \\
  NGC 300      & 6.9 & \phn 9.09 &  W  & NSC\tablenotemark{a} \\
  NGC 1249     & 5.9 & \phn 9.41 &  B  & NSC\tablenotemark{a} \\
  NGC 1494     & 7.0 & \phn 9.10 &  N  & NSC\tablenotemark{a} \\
  NGC 1518     & 7.7 & \phn 8.24 &  N  & NSC\tablenotemark{a} \\
  NGC 1559     & 5.9 & \phn 9.93 &  B  & NSC\tablenotemark{a} \\
  NGC 1744     & 6.5 & \phn 8.92 &  B  & NSC\tablenotemark{a} \\
  NGC 1796     & 5.0 & \phn 9.05 &  B  & NSC\tablenotemark{a} \\
  NGC 2427     & 7.6 & \phn 9.80 &  W  & NSC\tablenotemark{a} \\
  NGC 3621     & 6.8 & \phn 9.80 &  N  & NSC\tablenotemark{a} \\
  NGC 4504     & 6.1 & \phn 9.57 &  N  & NSC\tablenotemark{a} \\
  NGC 4781     & 6.8 & \phn 9.82 &  W  & NSC\tablenotemark{a} \\
  NGC 5264     & 9.3 & \phn 7.84 &  N  & NSC\tablenotemark{a} \\
  NGC 5334     & 5.0 & \phn 9.70 &  B  & NSC\tablenotemark{a} \\
  NGC 5713     & 4.1 &     10.35 &  B  & \ion{H}{2}/AGN\tablenotemark{d} \\
  NGC 6156     & 5.0 &     10.67 &  B  & AGN\tablenotemark{e} \\
  NGC 7456     & 5.9 & \phn 9.55 &  N  & unknown\tablenotemark{f} \\
  NGC 7713     & 6.9 & \phn 9.11 &  N  & NSC\tablenotemark{a} \\
  NGC 7793     & 7.0 & \phn 9.33 &  N  & NSC\tablenotemark{g} \\
  PGC 3853     & 6.9 & \phn 9.24 &  B  & NSC\tablenotemark{a} \\
  PGC 48179    & 8.9 & \nodata   &  W  & unknown\tablenotemark{f}
  \enddata
  
  \tablecomments{Col.~(1): Galaxy name. Col.~(2): Morphological type index.
    Col.~(3): Stellar mass. Col.~(4): Flag for the presence or absence of a bar:
    B = definitely barred; W = weakly barred; N = no bar or lens. Col.~(5):
    Presence of a nucleus and its physical nature from various references:
    active galactic nucleus (AGN), nuclear star cluster (NSC), and star-forming
    nucleus (\ion{H}{2}).}
  \tablenotetext{a}{\citealp{2014MNRAS+Georgiev}.}
  \tablenotetext{b}{\citealp{1983ApJ+Phillips}.}
  \tablenotetext{c}{Possible X-ray nucleus without optical counterpart on the
    $R$-band image; \citealp{2009ApJ+Desroches,2009ApJ+Zhang}.}
  \tablenotetext{d}{Starburst--AGN composite nucleus; \citealp{2010ApJ+Yuan}.}
  \tablenotetext{e}{The AGN may be saturated in the image;
    \citealp{2012ApJ+Alonso-Herrero}.}
  \tablenotetext{f}{The physical nature of the nucleus is unknown.}
  \tablenotetext{g}{\citealp{2002AJ+Boker}.}
\end{deluxetable}

\section{Notes on Individual Galaxies}
\label{sec:notes-indiv-galax}

\input{appendix_clean}

\bibliographystyle{aasjournal}
\bibliography{myref}
\end{CJK*}

\end{document}

%% file: figset.tex
\figsetstart
\figsetnum{2}
\figsettitle{Best-fit models of CGS disk galaxies.}

\figsetgrpstart
\figsetgrpnum{2.19}
\figsetgrptitle{ESO 506--G004}
\figsetplot{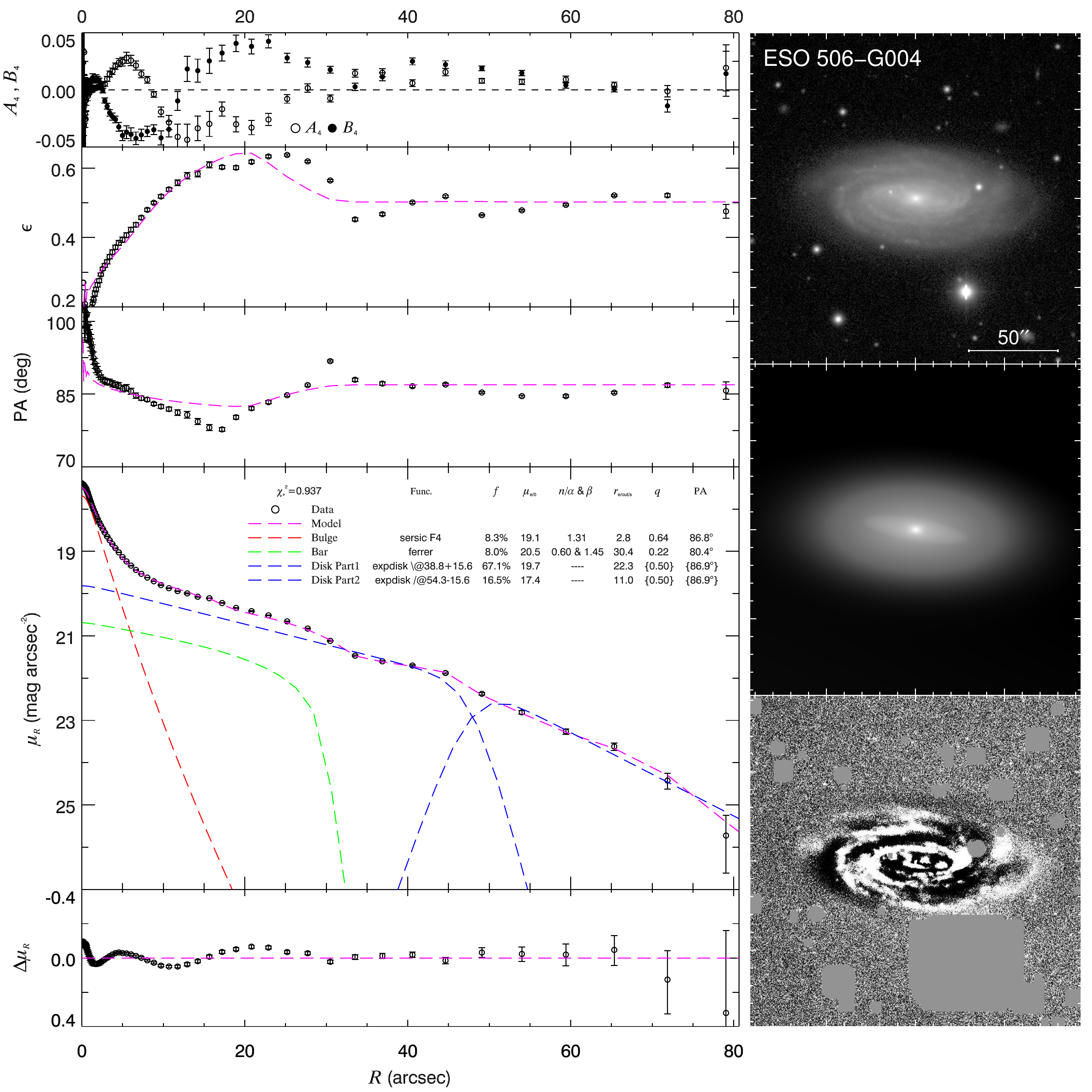}
\figsetgrpnote{Best-fit model of ESO 506--G004. The left panels display the isophotal analysis of the 2D image fitting. From top to bottom, the panels show the radial profiles of the fourth harmonic deviations from an ellipse ($A_{4}$ and $B_{4}$), ellipticity ($\epsilon$), position angle (PA), $R$-band surface brightness ($\mu_{R}$), and fitting residuals ($\bigtriangleup\mu_{R}$). The right panels display, from top to bottom, the grayscale $R$-band data image, the best-fit model image, and the residual images. The legends and explanatory text that gives details of each component follow the same convention as in the static version of this figure.}
\figsetgrpend

\figsetgrpstart
\figsetgrpnum{2.87}
\figsetgrptitle{NGC 1309}
\figsetplot{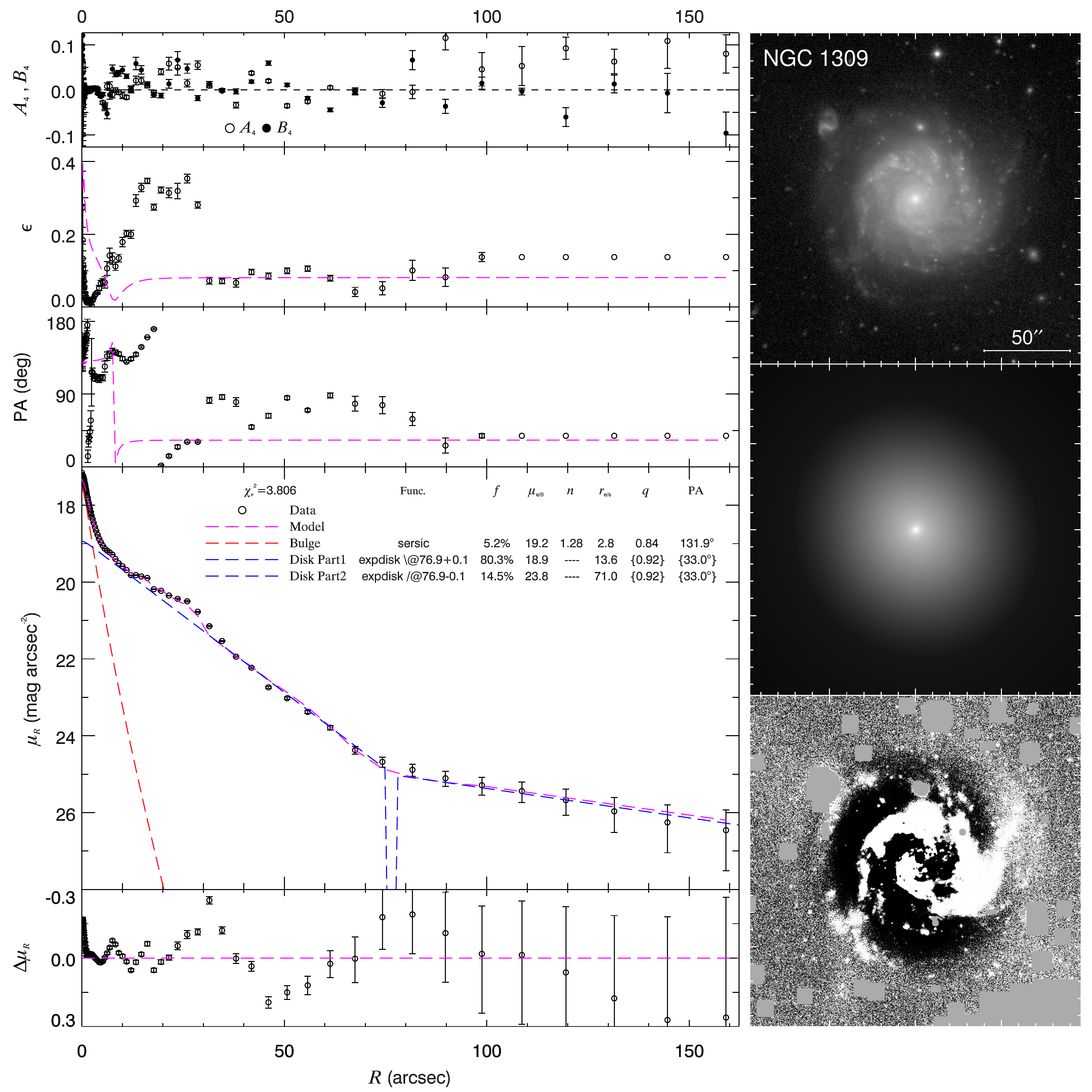}
\figsetgrpnote{Best-fit model of NGC 1309. The left panels display the isophotal analysis of the 2D image fitting. From top to bottom, the panels show the radial profiles of the fourth harmonic deviations from an ellipse ($A_{4}$ and $B_{4}$), ellipticity ($\epsilon$), position angle (PA), $R$-band surface brightness ($\mu_{R}$), and fitting residuals ($\bigtriangleup\mu_{R}$). The right panels display, from top to bottom, the grayscale $R$-band data image, the best-fit model image, and the residual images. The legends and explanatory text that gives details of each component follow the same convention as in the static version of this figure.}
\figsetgrpend

\figsetgrpstart
\figsetgrpnum{2.114}
\figsetgrptitle{NGC 1553}
\figsetplot{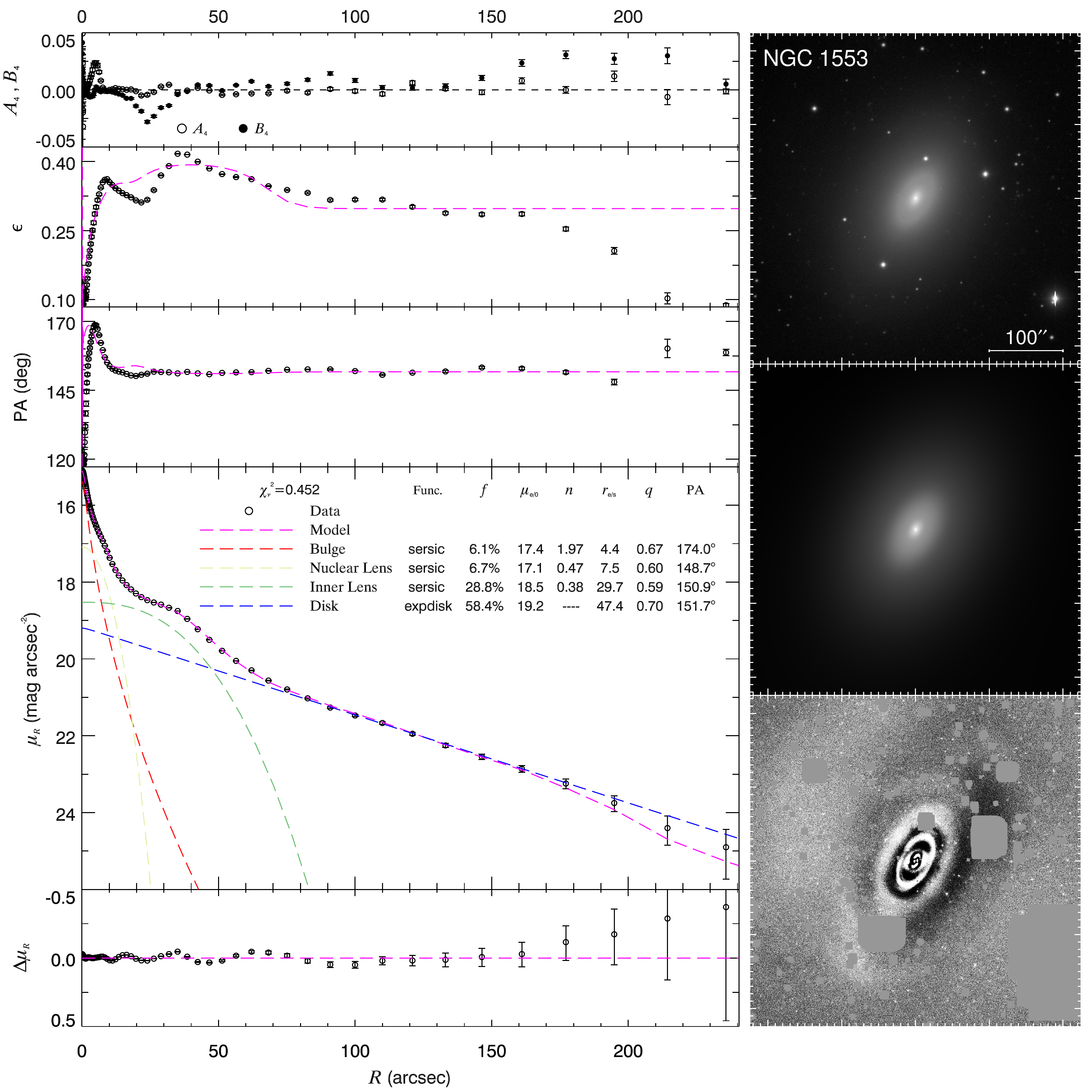}
\figsetgrpnote{Best-fit model of NGC 1553. The left panels display the isophotal analysis of the 2D image fitting. From top to bottom, the panels show the radial profiles of the fourth harmonic deviations from an ellipse ($A_{4}$ and $B_{4}$), ellipticity ($\epsilon$), position angle (PA), $R$-band surface brightness ($\mu_{R}$), and fitting residuals ($\bigtriangleup\mu_{R}$). The right panels display, from top to bottom, the grayscale $R$-band data image, the best-fit model image, and the residual images. The legends and explanatory text that gives details of each component follow the same convention as in the static version of this figure.}
\figsetgrpend

\figsetgrpstart
\figsetgrpnum{2.180}
\figsetgrptitle{NGC 3366}
\figsetplot{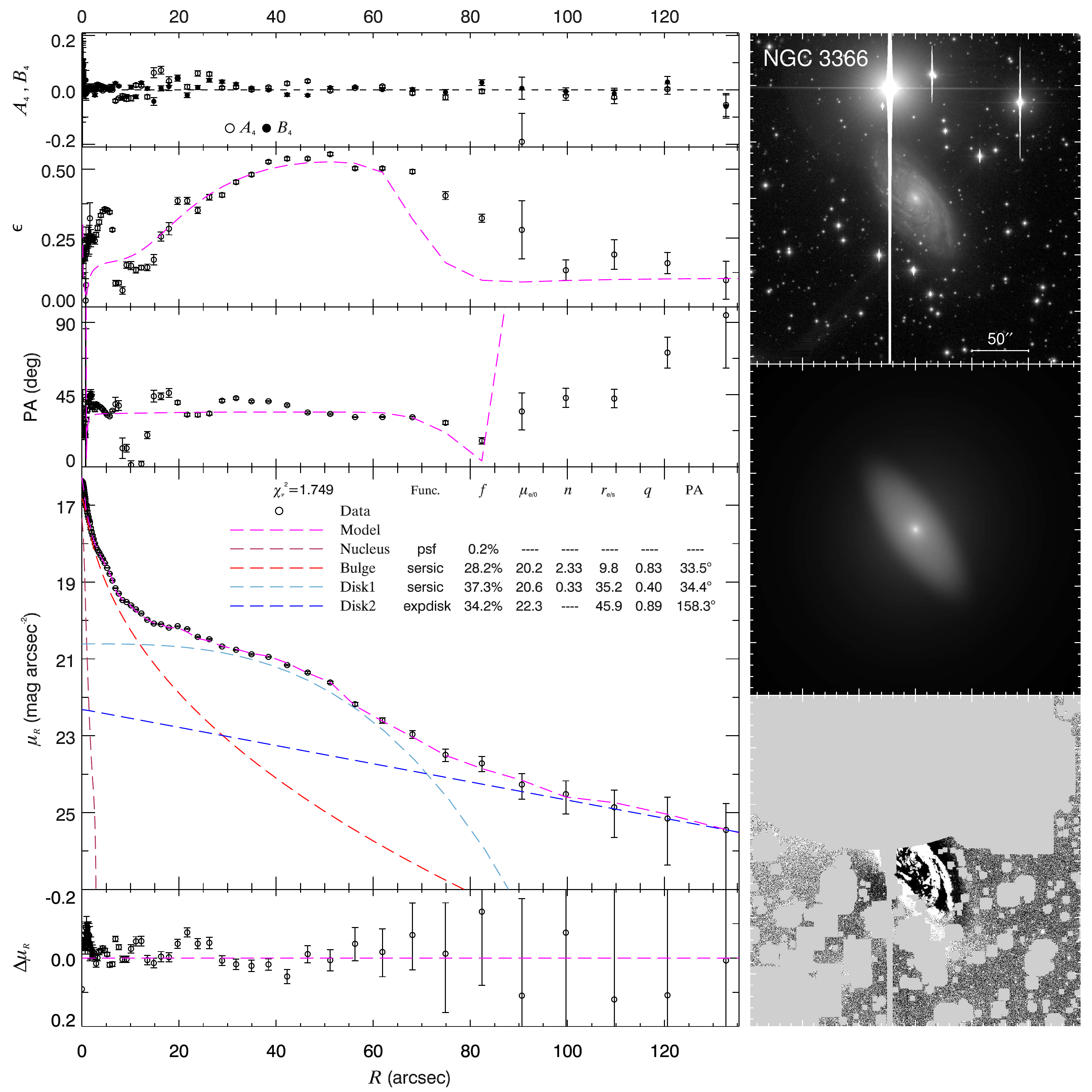}
\figsetgrpnote{Best-fit model of NGC 3366. The left panels display the isophotal analysis of the 2D image fitting. From top to bottom, the panels show the radial profiles of the fourth harmonic deviations from an ellipse ($A_{4}$ and $B_{4}$), ellipticity ($\epsilon$), position angle (PA), $R$-band surface brightness ($\mu_{R}$), and fitting residuals ($\bigtriangleup\mu_{R}$). The right panels display, from top to bottom, the grayscale $R$-band data image, the best-fit model image, and the residual images. The legends and explanatory text that gives details of each component follow the same convention as in the static version of this figure.}
\figsetgrpend

\figsetgrpstart
\figsetgrpnum{2.248}
\figsetgrptitle{NGC 5292}
\figsetplot{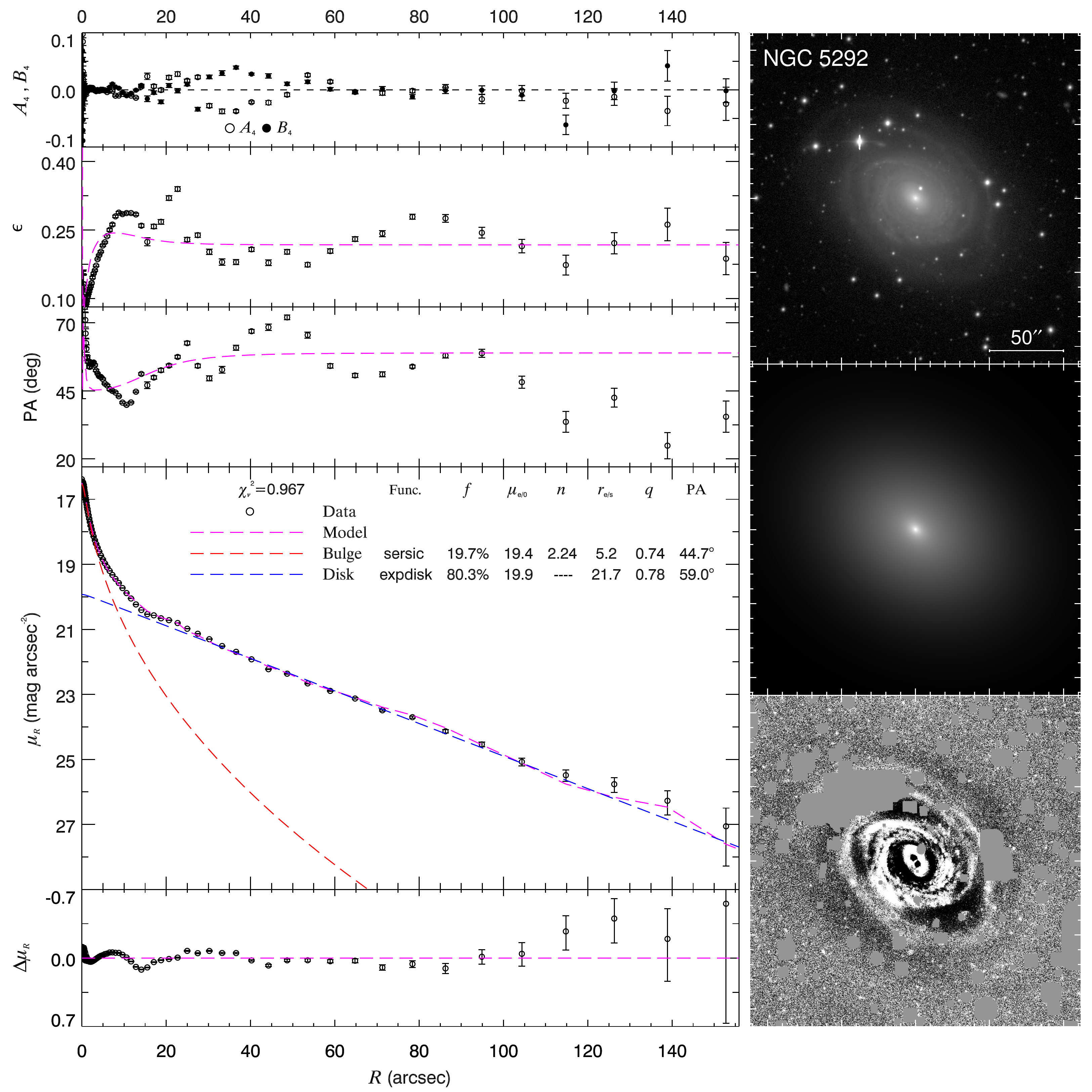}
\figsetgrpnote{Best-fit model of NGC 5292. The left panels display the isophotal analysis of the 2D image fitting. From top to bottom, the panels show the radial profiles of the fourth harmonic deviations from an ellipse ($A_{4}$ and $B_{4}$), ellipticity ($\epsilon$), position angle (PA), $R$-band surface brightness ($\mu_{R}$), and fitting residuals ($\bigtriangleup\mu_{R}$). The right panels display, from top to bottom, the grayscale $R$-band data image, the best-fit model image, and the residual images. The legends and explanatory text that gives details of each component follow the same convention as in the static version of this figure.}
\figsetgrpend

\figsetgrpstart
\figsetgrpnum{2.254}
\figsetgrptitle{NGC 5530}
\figsetplot{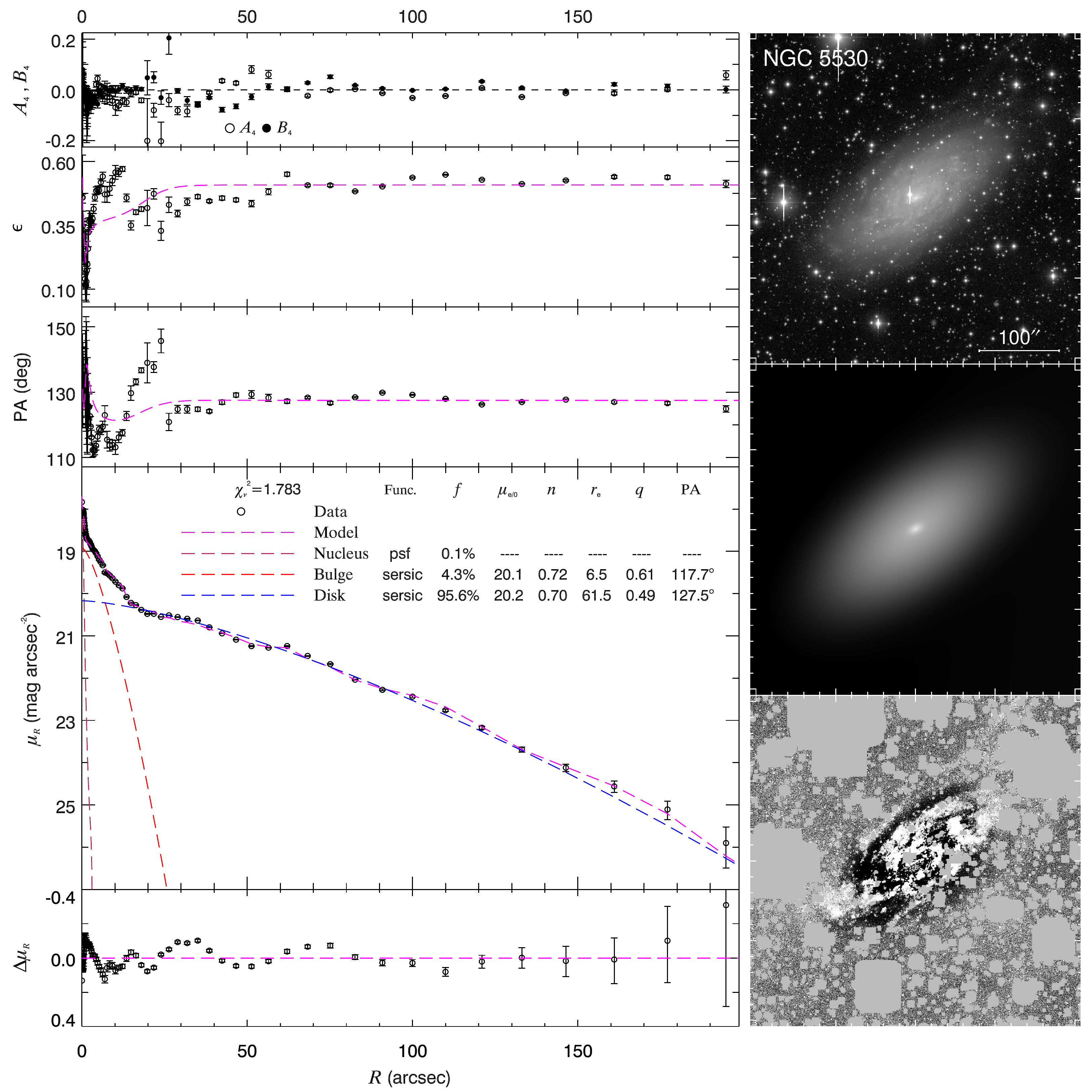}
\figsetgrpnote{Best-fit model of NGC 5530. The left panels display the isophotal analysis of the 2D image fitting. From top to bottom, the panels show the radial profiles of the fourth harmonic deviations from an ellipse ($A_{4}$ and $B_{4}$), ellipticity ($\epsilon$), position angle (PA), $R$-band surface brightness ($\mu_{R}$), and fitting residuals ($\bigtriangleup\mu_{R}$). The right panels display, from top to bottom, the grayscale $R$-band data image, the best-fit model image, and the residual images. The legends and explanatory text that gives details of each component follow the same convention as in the static version of this figure.}
\figsetgrpend

\figsetend

%% file: appendix_clean.tex

ESO 027--G001: The galaxy is barred. We model its disk break at the edge of the
spiral arms but ignore the disk break at $\sim140\arcsec$, which is treated
as an outer feature (e.g., outer lens and ring) when estimating bulge errors.

ESO 121--G026: We mask the dust lane on one side of the bar and model the disk
break at the inner ring.

ESO 137--G034: There are many foreground stars, including two saturated ones
near the bulge, that are carefully masked. We model the disk break at
$\sim40\arcsec$. The disk shows weak spiral arms. We mask the dust lane running
through the bulge.

ESO 138--G010: There are many foreground stars. The nucleus is an NSC
\citep{2014MNRAS+Georgiev}. The disk shows diffuse spiral arms.

ESO 186--G062: The galaxy has a high-$n$ bulge and a weak bar. We model its disk
break at the edge of the spiral arms.

ESO 213--G011: We model the disk break at the edge of the spiral arms.

ESO 221--G026: The galaxy is classified as an elliptical in both HyperLeda
\citep{2003A&A+Paturel} and the Third Reference Catalog of Bright Galaxies
(RC3; \citealp{1991Springer+de_Vaucouleurs}), but \citet{2013ApJ+Huang1}
discovered some substructures in it. The galaxy is likely to be an edge-on
system with a thin and a thick disk, which we provisionally model as a lens and
a disk in our decomposition (see Figure~2.7). Its bar/lens identification is
undecided (flagged as ``?'' in Table~\ref{tab:bul_param}).

ESO 221--G032: The galaxy has a compact bulge. We model the gentle disk break at
$\sim20\arcsec$. The circumnuclear dust lanes are masked during the fitting.

ESO 269--G057: The galaxy has an end-on bar whose $\alpha$ and $\beta$ are
fixed. We model the disk break/lens at the bar radius. The dust lanes around the
bulge are masked during the fitting.

ESO 271--G010: The galaxy has a weak bar whose $\alpha$ and $\beta$ need to be
fixed. We model the disk break at the edge of the spiral arms. A suspected weak
nucleus is neglected.

ESO 320--G026: We model the disk break at the edge of the spiral arms.

ESO 321--G025: The galaxy has a highly flattened bulge. The bar parameters have
to be fixed during the fitting. We model its disk break at the edge of the
spiral arms.

ESO 380--G001: The galaxy has a star-forming nucleus \citep{2010ApJ+Yuan}. The
bar parameters have to be fixed; otherwise, its length will be unrealistically
long. We model its disk break at the edge of the spiral arms. We mask the dust
lanes running through the bar. We ignore some outer features seen in the
residual image during the fitting and take them into account when estimating
the error budget for the bulge.

ESO 380--G006: We mask the dust lanes on the lens and near the bulge.

ESO 440--G011: The galaxy has a weak and flattened bulge.  The bar parameters
need to be fixed; otherwise, its length will be unrealistically long. We model
the disk break at the edge of the spiral arms.

ESO 442--G026: The galaxy is likely to be an edge-on system with a thin and a
thick disk, which we provisionally model as a lens and a disk in our
decomposition (see Figure~2.16). Its bar/lens identification is
undecided (flagged as ``?'' in Table~\ref{tab:bul_param}).

ESO 479--G004: The bar parameters have to be fixed otherwise its length will be
unrealistically long. We model the disk break at the edge of the spiral arms.

ESO 494--G026: The galaxy has a high-$n$ bulge and a bar. We model the disk
break at the edge of the spiral arms and mask dust lanes near the bulge.

ESO 506--G004: The galaxy is barred, and its bulge shows a weak X-shaped
feature. We model the disk break at the edge of the spiral arms.

ESO 507--G025: The galaxy is classified as an elliptical in HyperLeda but as an
S0 in RC3. We recognize a blue and dusty region around the galaxy center
($\sim30\arcsec$) and model it as an extra disk component. The dust lanes are
masked during the fitting.

ESO 582--G012: We model the disk break at the edge of the spiral arms and mask
dust lanes near the bulge.

IC 1953: We model the disk break at the edge of the spiral arms and mask dust
lanes across the bar and the bulge.

IC 1954: The bulge morphology is dominated by the nuclear bar. We model the disk
break at the edge of the spiral arms and mask dust lanes near the bulge.

IC 1993: We model the disk break at the edge of the spiral arms.

IC 2006: The galaxy is classified as an elliptical in HyperLeda but as an S0 in
RC3, and \citet{2013ApJ+Huang1} discovered some substructures in it. It has a
nuclear lens and an inner lens.

IC 2035: In addition to an extremely compact bulge, the galaxy hosts a short bar,
two lenses, and an underlying disk that exhibits different orientation. The
inner lens is difficult to model unless the outer lens is modeled
simultaneously.

IC 2051: The galaxy is barred. We model its disk break at the edge of the spiral
arms and mask dust lanes near the bulge.

IC 2056: The galaxy has an NSC \citep{2014MNRAS+Georgiev}. The bulge
exhibits nuclear spiral arms. We model the gentle disk break at the edge of the
spiral arms.

ICC 2367: We model the disk break at the bar radius.

IC 2522: The galaxy has a more flattened bulge than the disk.

IC 2537: We model the disk break at the edge of the spiral arms and mask the
dust lane near the bulge.

IC 2560: The nucleus appears to be bluer than the surrounding bulge and is
classified as Seyfert~2 by \citet{2010ApJ+Yuan}. The galaxy has a boxy/peanut
bulge. We model the disk break at the bar radius ($\sim50\arcsec$) but ignore
the outer disk break at the edge of the spiral arms ($\sim75\arcsec$), which is
treated as an outer feature when estimating bulge errors. We mask the dust lane
running through the bar.

IC 2627: The galaxy has a compact bulge.

IC 3253: The bulge has a large \sersic{} index ($n \approx 7$). The broken disk is
modeled with a \sersic{} function.

IC 4214: The bulge is well embedded in the lens or fat bar.

IC 4329: The galaxy is weakly barred. There is no disk break associated with the
weak bar. We need to fix some parameters of the bar component to ensure a
reasonable fitting.

IC 4444: There is a bright star close to the center. The galaxy has a lens. We
model the disk break at the edge of the spiral arms.

IC 4538: We model the disk break at the edge of the spiral arms and mask dust
lanes near the bulge.

IC 4618: The nucleus appears to be bluer than the surrounding bulge. The bent
bar is modeled with an $m=2$ bending mode. We model the disk break at the bar
radius. We mask the dust lanes near the bulge.

IC 4646: We model the gentle disk break at the edge of the spiral arms.

IC 4845: The galaxy shows a Type \Rmnum{3} disk profile.

IC 4901: The galaxy has a short bar, but it is not a nuclear bar, as it is
significantly longer than the bulge size. The galaxy has a two-disk
configuration. We ignore the suspected weak gentle disk break at the end of the
spiral arms.

IC 4946: The galaxy is barred and possibly has a boxy/peanut bulge. We mask the
dust lanes near the bulge.

IC 4991: A ring-like pattern shows up on the residual image. As we are not able
to identify a realistic ring structure and are unsure about its physical nature,
we attribute this pattern to artifacts and do not model it.

IC 5240: The galaxy has a boxy/peanut bulge, a strong bar, and a broken disk
with weak spiral arms. It is part of the training sample presented in
\citet{2017ApJ+Gao}. Here we show the decomposition results of the model that
includes all the above features (Model4 in their Table~9). Note that the
uncertainties are different from those presented in their Table~9, since we
include the model-induced uncertainties in this study.

IC 5267: The galaxy has an inner disk whose surface brightness profile is
reminiscent of a lens. The outer ring is visible on the residual image. The dust
lanes across the bulge is masked during the fitting.

IC 5273: The galaxy has a flattened bulge and a bar. Its disk shows a smooth
break and is lopsided.

IC 5325: We model the disk break at $\sim12\arcsec$ but ignore the outer disk
break at the edge of the spiral arms, which is treated as an outer feature when
estimating bulge errors.

IC 5332: The NSC \citep{2014MNRAS+Georgiev} manifest itself as an abrupt change
in color profile. The galaxy is angularly so large that simultaneously solving
for the sky level during the fitting is impossible. So we fix the sky level to the
value obtained via the direct approach. We use two disk components to account
for the plateau in the surface brightness profile at $\sim50\arcsec$ and the
underlying extended disk.

NGC 150: The galaxy has a starburst--AGN composite nucleus \citep{2010ApJ+Yuan}.
We model the disk break at the bar radius. 

NGC 151: The galaxy has an almost end-on bar. The broken disk is well-described
by a \sersic{} function.

NGC 210: The galaxy has a nuclear ring. The spiral arms/pseudo outer ring starts
at the ends of the lens/bar. The disk break at the edge of the spiral arms is
not modeled, as it is regarded as an outer feature. The dust lanes on the
lens/bar are not masked because they are far away from the bulge.

NGC 245: The parameters of the weak bar have to be fixed. We model the disk
break at the bar radius. The lopsided disk is modeled with an $m=1$ Fourier mode.

NGC 254: The galaxy has an inner lens and an outer ring. Inside $\sim5\arcsec$,
we find fine structures indicative of the presence of a nuclear ring and a
nuclear bar. This galaxy was used in \citet{2017ApJ+Gao} to illustrate that 
outer lenses/rings can be ignored for the purposes of bulge decomposition. Here
we present the full details of its decomposition, with the outer ring included
in the model.

NGC 255: The galaxy has a highly flattened bulge. We model the disk break at the bar
radius but ignore the outer disk break at the edge of the spiral arms, treating 
it as an outer feature when estimating bulge errors.

NGC 289: The galaxy is barred and has a two-disk configuration, with an inner
disk resembling a lens due to the tightly wound spiral arms and a diffuse
outer disk on which the arms unfold. We ignore the outer disk break at the edge
of the spiral arms ($\sim100\arcsec$), treatig it as an outer feature when
estimating bulge errors. We mask dust lanes near the bulge.

NGC 434: The galaxy has a bar whose $\alpha$ and $\beta$ are fixed. We model the
disk break that manifests as the spiral arms winding back to themselves. We mask
the dust lanes near the bulge and along the spiral arms.

NGC 578: The galaxy has a weak bar. We model the disk break at the edge of the
spiral arms. The disk is slightly lopsided. We mask central dust lanes during
the fitting.

NGC 584: The galaxy is classified as an elliptical in both HyperLeda and RC3,
but is recognized as an S0 in \citet{2013ApJ+Huang1}. It has a nuclear lens and
an inner lens.

NGC 613: The bar parameters are fixed to prevent it from being unrealistically
long. We model the inner disk break at the bar radius but ignore the outer disk
break ($\sim150\arcsec$), which is treated as an outer feature when estimating
bulge errors.

NGC 615: The galaxy has a lens. We model the disk break at the edge of the spiral
arms and mask the dust lanes near the bulge.

NGC 685: We ignore the suspected weak nucleus. The galaxy has a flattened bulge
and a bar. Its broken disk is modeled with a \sersic{} function. We mask the
dust lanes on the bar.

NGC 701: The galaxy has a highly flattened bulge. Its broken disk is modeled
with a \sersic{} function. We mask the dust lanes around the bulge.

NGC 782: The galaxy is barred whose $\alpha$ and $\beta$ are fixed. We model the
disk break at the inner ring.

NGC 895: We model the disk break at the edge of the spiral arms and mask the
spiral dust lanes approaching the bulge.

NGC 908: We model the disk break at the edge of the spiral arms and mask dust
lanes near the bulge.

NGC 936: The galaxy has a bar that is enclosed by an inner ring. Its structural
layout is similar to that of NGC~1533.

NGC 945: The galaxy has a weak bulge, a thin bar, and a broken disk with
prominent spiral arms. It is part of the training sample presented in
\citet{2017ApJ+Gao}. Here we show the decomposition results of the model that
includes all the above features (Model3 in their Table~11). Note that the
uncertainties are different from those presented in their Table~11, since we
include the model-induced uncertainties in this study.

NGC 986: We model the disk break at the bar radius as an inner lens/ring, but
ignore the outer disk break at the edge of the spiral arms and treat it as an
outer feature when estimating bulge errors. We mask the dust lanes running
through the bar and the bulge.

NGC 1022: The star-forming nucleus \citep{2010ApJ+Yuan} is distinctly
blue. The galaxy has a bar that is enclosed by an inner ring/lens. We mask the
circumnuclear dust lanes during the fitting.

NGC 1042: The nucleus is an NSC and an AGN \citep{2008ApJ+Shields,
  2014MNRAS+Georgiev}. We model the disk break at the edge of the spiral
arms. An extra disk component is needed to account for the bar-like pattern
produced by the spiral arms winding onto the bulge; otherwise, the bulge will be
unrealistically flattened.

NGC 1068: The galaxy has a Seyfert~2 nucleus \citep{1943ApJ+Seyfert}, a
flattened bulge that hosts a nuclear bar, and a prominent lens. The nucleus is
recently classified as a starburst--AGN composite nucleus by
\citet{2018MNRAS+DAgostino}. We mask the dust lanes near the center. The galaxy
is angularly so large that simultaneously solving the sky level during the
fitting is impossible. So we fix the sky level to the value obtained via the
direct approach.

NGC 1079: The galaxy has an inner ring/lens. We ignore the disk break at the
edge of the spiral arms, which is treated as an outer feature when estimating
bulge errors.

NGC 1084: The inner broken disk is modeled with a \sersic{} function.

NGC 1087: The galaxy has a flattened bulge. We ignore the disk break at the edge
of the spiral arms, because the overall disk is well-described by an exponential
function. We mask the circumnuclear dust lanes during the fitting.

NGC 1090: The galaxy is barred. We model its disk break at the edge of the
spiral arms and mask the circumnuclear dust lanes.

NGC 1097: The galaxy has a LINER/Seyfert~1 nucleus \citep{1997ApJ+Maiolino,
  2009ApJ+Ho} and a prominent nuclear star-forming ring well embedded in a
strong bar whose $\alpha$ and $\beta$ are fixed during the fitting. We ignore
the disk break at the edge of the spiral arms and treat it as an outer feature when
estimating bulge errors. The dust lanes along the bar are masked. The galaxy is
angularly so large that simultaneously solving the sky level during the fitting
is impossible. So we fix the sky level to the value obtained via the direct
approach.

NGC 1179: The galaxy has a short bar whose $\alpha$ and $\beta$ are fixed during
the fitting. We model the disk break at the bar radius but ignore the outer disk
break at the edge of the spiral arms, which is treated as an outer feature when
estimating bulge errors.

NGC 1187: We fix $\alpha$ and $\beta$ of the Ferrer bar during the fitting. We
model the disk break at the bar radius but ignore the outer disk break at the
edge of the spiral arms, treating it as an outer feature when estimating bulge
errors. We mask the dust lanes running across the bar.

NGC 1201: The galaxy contains an inner lens and an outer lens. But unlike normal
cases with two lenses of different sizes, in this case the inner lens fills the
outer lens in one dimension. Therefore, we also model the outer lens to avoid
potential bias of the bulge parameters. A possible outer ring is visible on the
residual image. There is a nuclear bar with a size of $\sim5\arcsec$ and a PA
$\approx$ $10\arcdeg$.

NGC 1232: The galaxy has a nuclear bar. We model the disk break at the edge of
the spiral arms. The galaxy is angularly so large that simultaneously solving
the sky level during the fitting is impossible. So we fix the sky level to the
value obtained via the direct approach.

NGC 1255: The galaxy has a flattened bulge. The disk is lopsided and breaks at
the edge of the spiral arms.

NGC 1291: The galaxy has a nuclear bar and a large-scale bar embedded in a
lens/ring. The dust lanes near the bulge are masked. The galaxy is angularly so
large that simultaneously solving the sky level during the fitting is
impossible. So we fix the sky level to the value obtained via the direct
approach.

NGC 1292: We model the disk break at the edge of the spiral arms.

NGC 1300: The galaxy has a prominent nuclear ring. We model the disk break at
the bar radius but ignore the outer disk break at the edge of the spiral arms,
treating it as an outer feature when estimating bulge error. The dust lanes
running through the bar are masked during the fitting.

NGC 1302: This is a barred galaxy with an inner ring and an outer ring.  This
galaxy was used in \citet{2017ApJ+Gao} to illustrate that the outer lenses/rings
can be ignored for the purposes of bulge decomposition. Here we present the full
details of its decomposition, with the outer ring included in the model.

NGC 1309: The galaxy has a Type \Rmnum{3} disk profile.

NGC 1317: The galaxy has a lens/weak bar. We find residual spiral patterns
outside the outer ring.

NGC 1326: The galaxy has a bar, a nuclear ring, an inner ring, and an outer
ring. It is part of the training sample presented in \citet{2017ApJ+Gao}. Here
we show the decomposition results that include the inner and outer ring, with
the nuclear ring unmasked (Model3 in their Table~8). Note that the uncertainties
are different from those presented in their Table~8, since we include the
model-induced uncertainties in this study.

NGC 1350: We model the disk break at the inner ring and ignore that at the outer
ring.

NGC 1353: We mask the dust lanes near the bulge during the fitting.

NGC 1357: The galaxy has a two-disk configuration, with the inner blue disk
showing prominent spiral arms and the outer red disk showing weak spiral
arms. It is part of the training sample presented in \citet{2017ApJ+Gao}. Here
we show the decomposition results that include all the above features (Model3 in
their Table~4). Note that the uncertainties are different from those presented
in their Table~4, since we include the model-induced uncertainties in this
study.

NGC 1365: The galaxy has a Seyfert~1.8 nucleus \citep{2010A&A+Veron-Cetty}, but 
it is recently classified as a starburst--AGN composite nucleus by
\citet{2018MNRAS+DAgostino}. We model the disk break at the edge of the spiral
arms. The dust lanes running through the bar are masked.

NGC 1367: The galaxy has a short bar whose $\alpha$ and $\beta$ are fixed during
the fitting. It has a two-disk configuration: a red inner disk and a blue outer disk.

NGC 1380: The galaxy is likely to be an edge-on system with a thin and a thick
disk, which we provisionally model as a lens and a disk in our decomposition
(see Figure~2.94). Its bar/lens identification is undecided (flagged
as ``?'' in Table~\ref{tab:bul_param}). The ``lens'' component is not perfectly
modeled by the \sersic{} function. The dust lane running through the bulge is
masked.

NGC 1385: We ignore the gentle disk break at $\sim60\arcsec$ and treat it as an
outer feature when estimating bulge errors. We mask the dust lanes to the north
of the bulge.

NGC 1386: The bulge is distinctly blue compared to the disk. The disk has a
Type~\Rmnum{2} profile. We tried to mask the majority of the dust lanes.

NGC 1387: The nuclear ring is readily recognizable in the residual pattern and
the color map. The galaxy is barred and its disk is broken at the bar radius.

NGC 1398: We model the disk break at the bar radius but ignore the outer disk
break at $\sim150\arcsec$, which is treated as an outer feature when estimating
bulge errors. The galaxy is angularly so large that simultaneously solving the
sky level during the fitting is impossible. So we fix the sky level to the value
obtained via the direct approach.

NGC 1400: The galaxy is classified as an elliptical in HyperLeda but as an S0 in
RC3. We recognize a lens at $\sim20\arcsec$. The dust lanes are masked during
the fitting.

NGC 1411: The galaxy has a nuclear lens and an inner lens. It is part of the
training sample presented in \citet{2017ApJ+Gao}. Here we show the decomposition
results of the model that includes the two lenses (Model3 in their
Table~2). Note that the uncertainties are different from those presented in
their Table~2, since we include the model-induced uncertainties in this study.

NGC 1415: The galaxy has a lens. We mask the dust lanes near the bulge during
the fitting.

NGC 1417: The galaxy has a high-$n$ bulge. We model the disk break at the edge of spiral arms.

NGC 1425: The galaxy has a high-$n$ bulge and a lens.

NGC 1433: The galaxy has a nuclear ring and a strong bar. We model the disk
break at the bar radius and mask the dust lanes running through the bar.

NGC 1436: We model the disk break at $\sim40\arcsec$. An extra disk component is
needed to account for the bar-like pattern produced by the spiral arms winding
onto the bulge; otherwise, the bulge orientation and ellipticity will be
incorrect.

NGC 1452: We model the disk break at the bar radius but ignore the outer disk
break at the edge of the spiral arms, which is treated as an outer feature when
estimating bulge errors. The disk shows weak spiral arms.

NGC 1493: The galaxy has an NSC \citep{2014MNRAS+Georgiev}. Recently it is
designated as an AGN candidate in X-rays, though it was classified as an
\ion{H}{2} nucleus in the optical \citep{2017ApJ+She1,2017ApJ+She2}. We fix
$\alpha$ and $\beta$ of the bar component during the fitting. We model the disk
break at the bar radius but ignore the outer disk break at the edge of the
spiral arms, treating it as an outer feature when estimating bulge errors. We
mask the dust lanes near the bulge.

NGC 1512: The galaxy and NGC~1510 form a starburst pair \citep{2006ApJS+Meurer}.
It has a starburst nucleus \citep{2011ApJ+Grier} and a nuclear ring. We model
the disk break at the bar radius but ignore the outer disk break at the edge of
the spiral arms, treating it as an outer feature when estimating bulge
errors. We mask the circumnuclear dust lanes and the dust lanes on the leading
edge of the bar.

NGC 1527: The galaxy has an inner lens and a weak outer lens.

NGC 1533: The galaxy is barred and its disk is broken roughly at the bar
radius. A ring-like pattern in the central 10\arcsec{} implies the presence of a
barlens--a face-on version of a boxy/peanut bulge. This galaxy is part of the
training sample presented in \citet{2017ApJ+Gao}. Here we show the decomposition
results of Model2 in their Table~7. Note that the uncertainties are different
from those presented in their Table~7, since we include the model-induced
uncertainties in this study.

NGC 1537: The galaxy is classified as an elliptical in HyperLeda but as a weakly
barred S0 in RC3. We recognize it as an S0 that has a nuclear lens and an inner
lens.

NGC 1543: The galaxy has a nuclear bar, a large-scale bar, an inner lens/ring,
and an outer ring.

NGC 1553: The galaxy has a nuclear lens and an inner lens/ring. Thus, its model
construction is similar to that of NGC~1411.

NGC 1566: The galaxy has an NSC \citep{2014MNRAS+Georgiev}. It is classified as
a Seyfert~1 nucleus in the optical \citep{2001ApJS+Sosa-Brito,
  2017MNRAS+da_Silva}. An extra disk component is needed to account for the
lens-like pattern produced by the spiral arms winding onto the bulge. We ignore
the disk break at the edge of the spiral arms and treat it as an outer feature
when estimating bulge errors. We mask circumnuclear dust lanes during the
fitting.

NGC 1574: There is a bright foreground star on the galaxy disk. The bar is
embedded in a lens. An outer ring is only visible on the residual image.

NGC 1617: We model the disk break at $\sim100\arcsec$. The weak spiral pattern
in the disk is visible on the residuals.

NGC 1637: The nucleus manifests itself as an abrupt change in the optical color
profile. The galaxy has a dominant point source in X-rays \citep{2009ApJ+Zhang,
  2013ApJ+Cisternas}, which suggests the presence of an AGN. We fix $\alpha$ and
$\beta$ of the bar component during the fitting. The disk is significantly
lopsided. We model the disk break at the edge of the spiral arms. We mask the
dust lanes near the bulge.

NGC 1640: The nucleus manifests itself as a abrupt change in the optical color
profile. The galaxy has a dominant point source in X-rays \citep{2009ApJ+Zhang,
  2013ApJ+Cisternas}, which suggests the presence of an AGN. We fix $\alpha$ and
$\beta$ of the bar component during the fitting. We model the disk break at the
bar radius but ignore the outer disk break at the edge of the spiral arms,
treating it as an outer feature when estimating bulge errors.

NGC 1667: The galaxy has a low-luminosity Seyfert~2 nucleus
\citep{1999ApJ+Barth,2010A&A+Veron-Cetty}, a nuclear bar, and a two-disk
configuration (inner blue disk and outer red disk).

NGC 1672: The galaxy has a dusty but overall blue bulge and a bar embedded in a
lens. We model the disk break at the edge of the spiral arms.

NGC 1688: The galaxy is reported to host an NSC by \citet{2002AJ+Carollo} and
\citet{2008ApJ+Seth}, but \citet{2014MNRAS+Georgiev} find no measurable
NSC. There is no X-ray detection in the nucleus \citep{2017ApJ+Foord}. The
galaxy has a highly flattened bulge. We fix $\alpha$ and $\beta$ of the bar
component during the fitting. The disk is significantly lopsided. We model the
disk break at the edge of the spiral arms. We mask the dust lanes near and on
the bulge.

NGC 1703: We model the disk break at the edge of the spiral arms.

NGC 1723: We model the disk break at the bar radius but ignore the outer disk
break at the edge of the spiral arms, treating it as an outer feature when
estimating bulge errors. We fix $\alpha$ and $\beta$ of the bar component during
the fitting.

NGC 1726: There are dust lanes near the galaxy center, and we mask them during
the fitting.

NGC 1784: We ignore both the gentle inner and outer disk breaks. The dust lanes
across the bar and the bulge are masked.

NGC 1792: We model the disk break at $\sim60\arcsec$.

NGC 1808: The galaxy has a Seyfert~2 nucleus according to
\citep{2011MNRAS+Brightman1}, while \citep{2010ApJ+Yuan} classified it as an
\ion{H}{2} nucleus. The dusty and star-forming bulge is embedded in a
lens/bar. We mask the dust lanes near the bulge.

NGC 1832: We ignore the gentle inner and outer disk breaks. We mask the dust
lanes near the bulge.

NGC 1947: There are many foreground stars throughout the image. We mask the dust
lanes across the bulge.

NGC 1954: The lens has a different orientation from that of the diffuse outer
disk.

NGC 1964: There is a bright star near the bulge. The galaxy has a dusty oval. We
mask the dust lanes near the bulge.

NGC 2082: There are many foreground stars throughout the image. The galaxy has a
weak NSC \citep{2002AJ+Carollo,2008ApJ+Seth} that does not affect the bulge
much. The galaxy has a bulge more flattened than the disk. We model the disk
break at the edge of the spiral arms.

NGC 2090: The galaxy has a two-disk layout: inner red disk/lens (perhaps due to
dust) and outer blue disk.

NGC 2139: The galaxy has an NSC \citep{2014MNRAS+Georgiev} and a highly
flattened, blue bulge. The lopsided disk shows a Type~\Rmnum{2} profile.

NGC 2196: The galaxy has an NSC \citep{2002AJ+Carollo} that is inactive
\citep{2004ApJ+Hunt}. We ignore the gentle disk break at $\sim50\arcsec$
because the overall disk is well-described by an exponential function.

NGC 2207: The galaxy is merging with IC~2163, but its overall morphology is
regular. The galaxy has a nuclear ring and a nuclear bar. We model the disk
break at the edge of the spiral arms. An extra disk component is needed to
account for the ring/plateau feature at $\sim30\arcsec$.

NGC 2217: The galaxy has a bar, and an inner and outer ring. The model includes
all these features, because it is difficult to achieve reasonable fits for the
bar and inner ring without the outer ring in the model. A nuclear ring with
a size of $\sim10\arcsec$ is visible in the residual pattern.

NGC 2223: The broken disk is modeled with a \sersic{} function. We fix $\alpha$
and $\beta$ of the bar component during the fitting.

NGC 2397: The galaxy has an NSC \citep{1997AJ+Carollo,2008ApJ+Seth}. We ignore
the disk break at $\sim30\arcsec$, because the overall disk is well-described by
an exponential function. We mask the dust lanes near the bulge.

NGC 2417: The suspected gentle disk break at $\sim40\arcsec$ is ignored.

NGC 2442: We model the disk break at the edge of the spiral arms. We mask the
dust lanes around the bulge.

NGC 2525: The galaxy has a photometrically distinct nucleus that is bluer than its
surroundings. We mask the dust lanes around the bulge and along the bar.

NGC 2559: The galaxy has a dusty and irregular bulge. We model the disk break at
the bar radius but ignore the outer disk break at the edge of the spiral arms,
treating it as an outer feature when estimating bulge errors. We mask the major
dust lanes along the bar and spiral arms.

NGC 2566: The galaxy has a blue star-forming bulge. We model the disk break at
the bar radius but ignore the outer disk break at the edge of the spiral arms,
treating it as an outer feature when estimating bulge errors. We mask the dust
lanes along the bar.

NGC 2640: The galaxy is weakly barred, and its disk is broken at the bar
radius. A large number of foreground stars are projected on top of the galaxy.

NGC 2695: The galaxy has an inner lens. 

NGC 2698: The galaxy is likely to be an edge-on system with a thin and a thick
disk, which we provisionally model as a lens and a disk in our decomposition
(see Figure~2.147). Its bar/lens identification is undecided (flagged
as ``?'' in Table~\ref{tab:bul_param}).

NGC 2708: The galaxy has an exponential lens/bar. We mask the dust lanes on the
lens and near the bulge.

NGC 2763: The galaxy has a very short bar whose size is comparable to that of the
bulge. Thus, we regard it as a nuclear bar.

NGC 2781: The galaxy has a nuclear ring, an inner lens/ring, and an outer
ring. We do not find any signature of a bar.

NGC 2784: The galaxy has an inner lens and outer lens. It is part of the
training sample presented in \citet{2017ApJ+Gao}. Here we show the decomposition
results of Model3 in their Table~3. Note that the uncertainties are different
from those presented in their Table~3, since we include the model-induced
uncertainties in this study.

NGC 2811: The inner empty region ($\la30\arcsec$) on the image and the peak in
the ellipticity profile suggest that there may be a bar, although modeling a bar is
difficult and uncertain due to the fact that the galaxy is highly inclined and the
bar is seen close to end-on. A composite disk model is constructed to make sure
that the disk break at the inner ring is properly taken into account.

NGC 2835: The galaxy has an NSC \citep{2014MNRAS+Georgiev} and a short bar.  We
model the disk break at the bar radius but ignore the outer disk break at the
edge of the spiral arms, treating it as an outer feature when estimating bulge
errors.  We mask the dust lanes near the bulge.

NGC 2848: We ignore the suspected weak disk break at the edge of the spiral
arms.

NGC 2889: The galaxy has a short and weak bar whose $\alpha$ and $\beta$ are
fixed. We mask the circumnuclear dust lanes. The broken disk is modeled with a
\sersic{} function. An extra disk component is included to account for the
diffuse outskirts.

NGC 2907: This is an almost edge-on galaxy with a thick disk that leads to an
underestimate of its inclination angle by \citet{2011ApJS+Ho}. We mask the dust
lanes running through the thin disk.

NGC 2935: The galaxy has a nuclear ring. Its bar is embedded in a lens/ring
structure. We model the disk break at the edge of the spiral arms.

NGC 2947: The galaxy has a distinctly blue nucleus. We model its disk break at
$\sim15\arcsec$.

NGC 2983: The galaxy is barred, and its disk is broken at the bar radius. Its
model construction is similar to that of NGC~1533.

NGC 3001: The galaxy has a distinctly blue nucleus of unknown nature
\citep{1986A&AS+Veron-Cetty}. An extra disk component must be included to
account for the tightly wound spiral arms that resemble a bar near the bulge;
otherwise, the bulge orientation and ellipticity will be incorrect. In addition,
the bulge orientation is constrained to be aligned with the disk. The size of
the original PSF image is not large enough, and we build an adequate one using
the IRAF task \texttt{psf}.

NGC 3038: An extra disk component is needed to account for the extra light
around the bulge; otherwise, the bulge will be unrealistically large. The dust
lane around the bulge is masked.

NGC 3052: The broken disk is modeled with a \sersic{} function.

NGC 3054: Significant sky gradient is present in the residual image. We model
the gentle disk break at the bar radius. We fix $\alpha$ and $\beta$ of the bar
component; otherwise, their values become unrealistic.

NGC 3056: The galaxy has an inner lens/ring. The residual pattern seems to
suggest the presence of a nuclear lens, but we do not find significant
signatures of substructures inside $\sim20\arcsec$ from inspection of its image
and isophotal analysis. So we do not pursue further refinements of the model.

NGC 3059: The galaxy has a highly flattened bulge. We mask the dust lanes along
the bar. The broken disk is modeled with a \sersic{} function. In addition, we
include an extra disk component to account for the diffuse outskirts.

NGC 3095: We mask the dust lanes along the bar.  We model the disk break at the
bar radius but ignore the outer disk break at the edge of the spiral arms,
treating it as an outer feature when estimating bulge errors.

NGC 3100: The galaxy has two lenses, but their configuration is unlike that of a
typical inner-outer lens configuration.  One lens fills the other in one
dimension; therefore, we model the two lenses together. There are dust lanes
near the bulge, which we mask during the fitting.

NGC 3108: This is an interesting case: a huge classical bulge is assembling a
diffuse disk around itself \citep{2008MNRAS+Hau}.

NGC 3124: The galaxy has a short and curved bar whose $\alpha$ and $\beta$ are
fixed. The broken disk is modeled with a \sersic{} function.

NGC 3145: We model the broken disk with a \sersic{} function.

NGC 3223: We model the disk break at $\sim90\arcsec$.

NGC 3261: We fix $\alpha$ and $\beta$ of the bar component during the
fitting. We model the disk break at the bar radius but ignore the outer disk
break at the edge of the spiral arms ($\sim60\arcsec$), treating it as an outer
feature when estimating bulge errors.

NGC 3271: Fortunately, we do not need to deal with the disk break associated with
the bar, as the bulge is well-embedded in the thick bar. The circular dust lane
at the galaxy center is masked during the fitting. We find fine structures that
suggest the presence of a nuclear bar roughly aligned with the large-scale bar.

NGC 3275: The bar parameters are fixed, or else the bar will be unrealistically
long. We model the disk break at the bar radius but ignore the outer disk break
at $\sim50\arcsec$, treating it as an outer feature when estimating bulge
errors.  We mask the dust lane across the bar.

NGC 3281: The galaxy is classified as Seyfert~2 in the optical
\citep{2010A&A+Veron-Cetty}, but we find no sign of a nucleus, which is probably
obscured by the dust. The galaxy is well-described by a \sersic{} bulge and an
exponential disk. We mask the dust lanes running through the center.

NGC 3313: The galaxy has a prominent nuclear star-forming ring. We model the
disk break at the bar radius. We carefully mask the bright star near the bulge.

NGC 3318: The galaxy has a compact bulge. We fix $\alpha$ and $\beta$ of the
weak bar during the fitting. We model the disk break at the bar radius.

NGC 3358: The galaxy has an inner lens. The outer disk break/ring is also
modeled, or else the bulge will be underestimated.

NGC 3366: The galaxy hosts an inactive nucleus \citep{2008A&A+Siebenmorgen}. The
broken disk is modeled with a \sersic{} function. We include an extra disk
component to account for the diffuse outskirts.

NGC 3450: We fix $\alpha$ and $\beta$ of the bar component during the
fitting. We model the disk breaks at the bar radius and at the edge of the
spiral arms.

NGC 3513: There is noticeable residual light at the center of the galaxy. Since
\citet{2014MNRAS+Georgiev} did not find a measurable NSC, we do not include a
PSF component to account for the residuals. The galaxy has a highly flattened
bulge. We fix $\alpha$ and $\beta$ of the bar component during the fitting. We
model the disk break at the bar radius but ignore the outer disk break at the
edge of the spiral arms ($\sim60\arcsec$), treating it as an outer feature when
estimating bulge errors. We mask the dust lanes near the bulge.

NGC 3521: The galaxy has an emission-line nucleus classified as
\ion{H}{2} or LINER \citep{1997ApJS+Ho}. \citet{2014MNRAS+Georgiev} did not find
any measurable NSC. We use two exponential components with different
ellipticities to model the disk.

NGC 3568: We use two components with slightly different orientations and
ellipticities to model the disk. We mask the dust lanes near the bulge.

NGC 3660: We model the disk break at the bar radius but ignore the outer disk
break at the edge of the spiral arms, treating it as an outer feature when
estimating bulge errors.  We mask the dust lanes running through the bar and the
bulge.

NGC 3672: We model the disk break at the edge of the spiral arms.

NGC 3673: The galaxy has a distinctly blue nucleus and a weakly boxy bulge.  We
model the disk break at the bar radius but ignore the outer disk break at the
edge of the spiral arms, treating it as an outer feature when estimating bulge
errors. We mask the dust lanes near the bulge.

NGC 3763: The broken disk is modeled with a \sersic{} function. Some of the bar
parameters are fixed. We mask the dust lanes around the bulge.

NGC 3783: The galaxy has a Seyfert~1/1.5 nucleus \citep{2010A&A+Veron-Cetty,
  2010ApJ+Yuan}. We model the disk break at the bar radius and the
anti-truncation at $\sim60\arcsec$.

NGC 3882: There are many foreground stars throughout the image. We model the
disk break at the bar radius. We mask the dust lanes near the bulge.

NGC 3885: The galaxy has a lens. We mask all the major dust lanes during the
fitting.

NGC 3887: The galaxy has a distinctly blue nucleus. We model the disk break at
the bar radius. We model the spiral dust lanes approaching the galaxy center.

NGC 3892: This barred galaxy has an inner ring and an outer ring. In addition,
we need to include a compact nucleus, which is modeled with a PSF component, or
else the \sersic{} index of the bulge would be unrealistically large.

NGC 3904: The galaxy is classified as an elliptical in both HyperLeda and RC3,
but is recognized as a possible S0 in \citet{2013ApJ+Huang1}. It has two lenses,
one filling the other in one dimension. We model both lenses simultaneously.

NGC 3955: The galaxy hosts a weak nucleus of unknown nature \citep{2010ApJ+Yuan}
that appears to be abruptly bluer than its surrounding. We mask dust-obscured
regions within $\sim40\arcsec$. The galaxy has a two-disk configuration: an
inner dusty but blue disk and an outer smooth one.

NGC 3981: We fix $\alpha$ and $\beta$ of the bar component during the
fitting. The galaxy has a two-disk configuration: an inner disk with sharply
truncated spiral arms and an outer diffuse one. We mask the dust lanes near the
bulge. Note that there are significant residuals of unknown origin at the galaxy
center.

NGC 4024: The galaxy is barred, and its disk break at the bar radius is
weak. Its model construction is similar to that of NGC~1533.

NGC 4027: The galaxy has a flattened bulge. We model the inner disk break but
ignore the outer one, which is treated as an outer feature when estimating bulge
errors. The disk is significantly lopsided. We mask the dust lanes around the
bulge.

NGC 4030: \citet{2014MNRAS+Georgiev} found an NSC at the galaxy center, but we
find no sign of an unresolved point source on the CGS image. We need to include
an extra disk component in the model; otherwise, the bulge will be
unrealistically large.

NGC 4033: The galaxy is classified as an elliptical in both HyperLeda and RC3,
but is recognized as a possible S0 in \citet{2013ApJ+Huang1}. It has a nuclear
lens.

NGC 4050: We model the disk break at the bar radius but ignore the outer disk
break at $100\arcsec$, treating it as an outer feature when estimating bulge
errors. We mask the dust lanes around the bulge.

NGC 4094: We model the disk break at the edge of the spiral arms. We mask the
dust lanes around the bulge.

NGC 4304: We model the disk break at the bar radius but ignore the outer disk
break at the edge of the spiral arms, treating it as an outer feature when
estimating bulge errors.  We mask the dust lanes running through the bar and the
bulge.

NGC 4373A: The galaxy is likely to be an edge-on system with a thin and a thick
disk, which we provisionally model as a lens and a disk in our decomposition
(see Figure~2.203). Its bar/lens identification is undecided (flagged
as ``?''  in Table~\ref{tab:bul_param}). The dust lane running through the bulge
is masked during the fitting.

NGC 4462: We model the disk break at the bar radius but ignore the outer disk
break at the edge of the spiral arms, treating it as an outer feature when
estimating bulge errors.  We mask the dust lanes running through the bar and
around the bulge.

NGC 4487: The galaxy has an NSC \citep{2014MNRAS+Georgiev} and a flattened
bulge. We model the disk break at the edge of the spiral arms. We mask the dust
lanes near the bulge.

NGC 4546: The galaxy is likely to be an edge-on system with a thin and a thick
disk, which we provisionally model as a lens and a disk in our decomposition
(see Figure~2.206). Its bar/lens identification is undecided (flagged
as ``?'' in Table~\ref{tab:bul_param}).

NGC 4593: The galaxy has a Seyfert~1 nucleus \citep{2010A&A+Veron-Cetty,
  2010ApJ+Yuan}, a bar embedded in a lens/ring component, and a disk break at
the edge of the spiral arms. The dust lanes around the bulge are masked.

NGC 4594: The Sombrero galaxy is an edge-on galaxy with a thick disk that
led to an underestimate of its inclination angle by \citet{2011ApJS+Ho}. We mask
the major dust lane running through the thin disk.

NGC 4603: We model the broken disk with a \sersic{} function.

NGC 4632: The galaxy hosts an \ion{H}{2} nucleus \citep{2007MNRAS+Decarli,
  2013A&A+Gavazzi}. We model the disk break at $\sim60\arcsec$. We mask the dust
lanes near the bulge.

NGC 4650: We model the gentle disk break at the bar radius and mask the dust lanes
near the bulge.

NGC 4653: We ignore the gentle disk break at $\sim90\arcsec$.

NGC 4684: We attribute the lens-like structure with a size of $\sim20\arcsec$ as
the bulge. Otherwise, the galaxy would have $B/T = 0$. The compact nucleus is
modeled as a PSF component. The central dust lane is masked during the fitting.

NGC 4691: The nucleus is classified as an \ion{H}{2} nucleus in the optical
\citep{2010ApJ+Yuan}, but there is no evidence for an obscured AGN in the X-rays
\citep{2003MNRAS+Maiolino}. However, \citet{2010A&A+Veron-Cetty} classified it
as Seyfert~1 nucleus. The galaxy has a highly flattened bulge. We model the disk
break at the bar radius. We mask the dust lanes along the bar and on the bulge.

NGC 4697: The galaxy is classified as an elliptical in both HyperLeda and RC3,
but is recognized as an S0 in \citet{2013ApJ+Huang1}. The galaxy is likely to be
an edge-on system with a thin and a thick disk, which we provisionally model as
a lens and a disk in our decomposition (see Figure~2.215). Its bar/lens
identification is undecided (flagged as ``?'' in Table~\ref{tab:bul_param}).

NGC 4699: The galaxy has a nuclear bar and two lenses.

NGC 4727: The galaxy has a short bar whose $\alpha$ and $\beta$ are fixed during
the fitting. It has a two-disk configuration: an inner disk with spiral arms and
an outer diffuse disk without discernible spiral patterns.

NGC 4731: The galaxy has a highly flattened bulge.  We model the disk break at
the bar radius but ignore the outer disk break at the edge of the spiral arms,
treating it as an outer feature when estimating bulge errors. We mask the dust
lanes near the bulge.

NGC 4802: The galaxy has a dusty but overall blue bulge, which is indicative of
ongoing star formation. In addition, we recognize a nuclear lens and an inner
lens. We mask the dust lanes around the bulge. The compact nucleus is modeled
using a PSF component.

NGC 4825: The galaxy is classified as an elliptical in HyperLeda but as an S0 in
RC3. The central dust lane running through the bulge is masked during the
fitting.

NGC 4856: The galaxy is relatively edge-on, but its bar is still readily
recognizable. Its disk is broken at the bar radius.

NGC 4899: We model the disk break at the edge of the spiral arms and mask the
dust lanes on the bulge.

NGC 4902: We model the disk break at the bar radius but ignore the outer disk
break at the edge of the spiral arms, treating it as an outer feature when
estimating bulge errors. We mask the dust lanes near the bulge during the
fitting.

NGC 4930: We fix $\alpha$ and $\beta$ of the bar component; otherwise, it will
be unrealistically long. We model the disk break at the bar radius.

NGC 4939: The galaxy has tightly wound spiral arms. We ignore the possible
gentle disk break beyond $\sim 170\arcsec$.

NGC 4941: The galaxy has a Seyfert~2 nucleus \citep{1993ApJS+Rush,
  2010A&A+Veron-Cetty}, but whether we include a nucleus in the model or not
does not affect the bulge parameters much. The galaxy probably hosts a nuclear
bar. The bulge is embedded in a lens.

NGC 4947: The galaxy has an \ion{H}{2} nucleus \citep{1986A&AS+Veron-Cetty}. The
bulge orientation is constrained to be the same as that of the disk, or else
the bulge will turn to fit the spiral arms winding onto itself. We model the
broken disk with a \sersic{} function.  We find some positive residuals beyond
$\sim 60\arcsec$ but do not include an extra component to account for them. We
treat them as outer features when estimating bulge errors. We mask the dust
lanes near the bulge during the fitting.

NGC 4965: The galaxy has a distinctly blue nucleus of unknown nature. We model
the disk break at the edge of the spiral arms. Note that there is a bright blob
to the north-east of the galaxy, which causes noticeable residuals in the 
background.

NGC 4981: We fix $\alpha$ and $\beta$ of the weak bar during the fitting.  We
model the disk break at the bar radius but ignore the outer disk break at the
edge of the spiral arms, treating it as an outer feature when estimating bulge
errors. We mask the dust lanes on the bulge and the bar.

NGC 4984: The galaxy has an inner lens and outer ring. The bulge is
distinctly blue compared with the lens and the disk. This galaxy was used in
\citet{2017ApJ+Gao} to illustrate that outer lenses/rings can be ignored for
the purposes of bulge decomposition. Here we present the full details of its
decomposition, with the outer ring included in the model.

NGC 4995: The galaxy has a distinctly blue nucleus that is classified as a
composite AGN/\ion{H}{2} nucleus by \citet{1994ApJ+Giuricin} but as an
\ion{H}{2} nucleus by \citet{1993ApJS+Rush}. We model the disk break at
$\sim30\arcsec$.

NGC 5026: The galaxy has a bar that is enclosed by an inner ring. An outer ring
is visible on the residual image.

NGC 5042: The galaxy has a distinctly blue nucleus of unknown nature. We ignore
the gentle disk break at $\sim140\arcsec$. We mask the circumnuclear dust lanes
during the fitting.

NGC 5054: The galaxy has an NSC \citep{2014MNRAS+Georgiev} and a nuclear
ring. We mask the dust lanes near the bulge.

NGC 5068: The galaxy has an NSC \citep{2014MNRAS+Georgiev} and a highly
flattened bulge. We model the disk break at the bar radius and mask the dust
lanes near the bar during the fitting.

NGC 5078: The is an edge-on galaxy with a thick disk that leads to an
underestimate of its inclination angle by \citet{2011ApJS+Ho}. We mask the
prominent dust lane running through the thin disk.

NGC 5101: We model the disk break at the bar radius. We fix $\alpha$ and $\beta$
of the bar component, or else the bar will be unrealistically long.

NGC 5121: The galaxy has a lens. We model the disk break at the edge of the
spiral arms.

NGC 5134: The galaxy has a distinctly red and inactive nucleus
\citep{2014A&A+Koulouridis}. The galaxy has a lens. We mask the dust lanes near
the bulge.

NGC 5135: The bar is embedded in a lens. We ignore the disk break at the edge of
the spiral arms and treat it as an outer feature when estimating bulge errors.

NGC 5156: We model the disk break at the bar radius but ignore the outer disk
break at $\sim60\arcsec$, treating it as an outer features when estimating bulge
errors. We mask the dust lanes on the bulge and the bar.

NGC 5188: The center is heavily dust-obscured. The galaxy has a lens. We ignore
the disk break at $\sim60\arcsec$ and treat it as an outer feature when estimating
bulge errors.

NGC 5247: The galaxy has a dusty but overall blue bulge. We model the disk break
at the edge of the spiral arms. Meanwhile, we fix the scale length of the inner
part of the disk; otherwise, it will be unrealistically long. We mask the dust
lanes on and near the bulge.

NGC 5253: The galaxy has an NSC \citep{2016ApJ+Smith} and a starburst bulge. We
mask the dust lanes on the bulge. The peculiar pattern in the sky background
introduces large sky measurement error.

NGC 5254: We model the broken disk with a \sersic{} function.

NGC 5266: We mask the central circular dust lanes along the minor axis of the
galaxy.

NGC 5292: The galaxy is well-described by a \sersic{} bulge and an exponential
disk.

NGC 5324: The galaxy has a nuclear ring. We model the broken disk with a
\sersic{} function.

NGC 5333: The galaxy has a nuclear lens and an inner lens.

NGC 5339: The galaxy has a distinctly blue nucleus of unknown nature. We model
the disk break at the bar radius. We mask the dust lanes along the bar and around
the nucleus.

NGC 5468: The galaxy has a compact bulge. An extra disk component is needed to
account for the lens-like pattern produced by the spiral arms winding onto the
bulge. The broken disk is modeled with a \sersic{} function.

NGC 5483: The galaxy has a nucleus of unknown nature. We ignore the positive residuals in the outskirts and treat them as outer features when estimating bulge errors.

NGC 5530: The galaxy has an NSC \citep{2014MNRAS+Georgiev}. There is a bright
saturated star near the center. We model the broken disk with a \sersic{}
function.

NGC 5556: The galaxy has an NSC \citep{2014MNRAS+Georgiev} and a flattened
bulge. We model the disk break at the edge of the spiral arms. We mask several
dust-attenuated regions around the bulge.

NGC 5597: The galaxy has an \ion{H}{2} nucleus \citep{2004ApJ+Hunt} and a
flattened bulge. We model the disk break at the edge of the spiral arms. We mask
several dust-attenuated regions around the bulge and on the bar.

NGC 5643: The galaxy has a Seyfert nucleus \citep{1983ApJ+Phillips}. We model
the disk break at the bar radius but ignore the outer disk break at
$\sim60\arcsec$, treating it as an outer feature when estimating bulge
errors. We mask the dust lanes along the bar and around the bulge.

NGC 5688: The galaxy has an almost end-on bar, whose $\alpha$ and $\beta$ are
fixed. We model the broken disk with a \sersic{} function.

NGC 5728: The galaxy has a nuclear ring. We model the disk break at the bar
radius but ignore the outer disk break at the edge of the spiral arms, treating
it as an outer feature when estimating bulge errors.

NGC 5786: The galaxy has a blue bulge with a large \sersic{} index and a short
bar. We model the disk break at the edge of the spiral arms. We mask the dust
lanes near the bulge. The image shows significant sky gradient introduced by a
saturated star.

NGC 5833: The galaxy has a distinctly blue nucleus that is inactive
\citep{2002ApJ+Greenhill}. We ignore the gentle disk break at $\sim 60\arcsec$
and treat it as an outer feature when estimating bulge errors. We mask the dust
lanes near the bulge.

NGC 5861: The galaxy has a dusty bulge. We model the broken disk with a
\sersic{} function. We mask the dust lanes on the bulge.

NGC 5885: The galaxy has a blue compact bulge and a short bar whose $\alpha$ and
$\beta$ are fixed. We ignore the gentle disk break at $\sim 100\arcsec$ and
treat it as an outer feature when estimating bulge errors. We mask the dust lanes
near the bulge and along the bar.

NGC 5938: There are many foreground stars throughout the image. We model the
disk break at the bar radius but ignore the outer disk break at the edge of the
spiral arms, treating it as an outer feature when estimating bulge errors. We
mask the dust lanes around the bulge.

NGC 5967: The galaxy has an \ion{H}{2} nucleus \citep{1986A&AS+Veron-Cetty} and
a weak bar whose parameters have to be fixed. We model the broken disk with a
\sersic{} function.

NGC 6118: The galaxy has a flattened bulge and a broken spiral disk
well-described by a \sersic{} function. It is part of the training sample
presented in \citet{2017ApJ+Gao}. Here we show the decomposition results that
include all the above features (Model3 in their Table~6). Note that the
uncertainties are different from those presented in their Table~6, since we
include the model-induced uncertainties in this study.

NGC 6215: There are many foreground stars throughout the image. We model the
disk break at $\sim 40\arcsec$. We mask the dust lanes on and around the bulge.

NGC 6221: The galaxy has a Seyfert~2 nucleus \citep{1981A&A+Veron}. We model the
disk break at the edge of the spiral arms. An extra disk component is needed to
account for the bar-like pattern produced by the spiral arms winding onto the
bulge; otherwise, the bulge orientation will be incorrect. We mask the dust
lanes on and around the bulge.

NGC 6300: The galaxy has a Seyfert~2 nucleus \citep{1983ApJ+Phillips}, which was
designated later as a changing-look AGN \citep{2003MNRAS+Matt}. We model the
disk break at the bar radius but ignore the outer disk break at the edge of the
spiral arms, treating it as an outer feature when estimating bulge errors.  We
mask the dust lanes around the bulge.

NGC 6392: The galaxy has a short and weak bar whose $\alpha$ and $\beta$ have to
be fixed. It has a two-disk configuration: an inner blue disk with spiral arms
and an outer smooth red disk.

NGC 6492: The galaxy has a lens. We model the disk break at the edge of the
spiral arms.

NGC 6673: The galaxy is classified as an elliptical in HyperLeda but as an S0 in
RC3, and is recognized as a possible S0 in \citet{2013ApJ+Huang1}. It has a
nuclear lens and an inner lens.

NGC 6684: The galaxy has a bar, an inner ring, and an outer ring/lens. A nuclear
bar embedded in the bulge is roughly perpendicular to the large-scale bar.

NGC 6699: We model the broken disk with a \sersic{} function.

NGC 6744: The galaxy has a LINER nucleus \citep{1997AJ+Vaceli}.  We model the
disk break at the bar radius but ignore the outer disk break at the edge of the
spiral arms, treating it as an outer feature when estimating bulge errors.  The
galaxy is angularly so large that simultaneously solving the sky level during
the fitting is impossible. So we fix the sky level to the value obtained via the
direct approach. We mask the dust lanes around the bulge.

NGC 6753: The bulge is embedded in a nuclear lens and an inner lens.

NGC 6754: We model the disk break at the edge of the spiral arms. An extra disk
component is needed to account for the bar-like pattern produced by the spiral
arms winding onto the bulge, or else the bulge orientation and ellipticity will
be incorrect. The disk is slightly lopsided. We mask the dust lanes around the
bulge.

NGC 6782: The galaxy has a nuclear ring and a nuclear bar. We fix $\alpha$ and
$\beta$ of the bar component during the fitting. We model the disk break at the 
bar radius.

NGC 6788: The galaxy is well-described by a \sersic{} bulge and an exponential
disk, except for the central positive residuals of unknown physical nature.

NGC 6810: The galaxy has a distinctly blue nucleus with ambiguous
classifications: Seyfert~2 nucleus \citep{1990AJ+Kirhakos}, \ion{H}{2} nucleus
\citep{1986A&AS+Veron-Cetty,2007MNRAS+Strickland,2008MNRAS+Brightman,
  2013ApJS+Videla,2014MNRAS+Asmus}, and \ion{H}{2}/AGN composite nucleus
\citep{2011MNRAS+Brightman2,2010ApJ+Yuan}. This galaxy is actually edge-on, but
its inclination angle is underestimated due to the presence of the thick
disk. In addition, a lens-like component is found around the bulge. We mask the
prominent dust lanes throughout the galaxy.

NGC 6814: The galaxy has a Seyfert~1.5 nucleus \citep{2010A&A+Veron-Cetty}. We
model the disk break at the bar radius.

NGC 6893: The galaxy has an inner lens and an outer lens. This galaxy was used
in \citet{2017ApJ+Gao} to illustrate that the outer lenses/rings can be ignored
for the purposes of bulge decomposition. Here we present the full details of its
decomposition, with the outer lens included in the model.

NGC 6902: The bulge is surrounded by a ring/lens feature (Disk1 in Figure 2.282).

NGC 6907: The galaxy has a nuclear bar. We model the disk break at the edge of
the spiral arms. The dust lanes near the bulge are masked.

NGC 6923: The galaxy has a short bar/lens. We model the disk break at the edge
of the spiral arms. The dust lanes near the bulge are masked.

NGC 6935: The galaxy has a high-$n$ bulge. We model the disk break at
$\sim20\arcsec$.

NGC 6942: The galaxy is barred and shows a disk break at $\sim50\arcsec$. Spiral
patterns are visible on the residual image, but they are quite weak and can be
ignored.

NGC 6943: The galaxy has a weak bar. We model the disk break at $\sim50\arcsec$.

NGC 7038: We model the disk break at the edge of spiral arms ($\sim80\arcsec$).

NGC 7049: The galaxy has a lens. The circular dust lane around the bulge is
masked during the fitting.

NGC 7059: We model the disk break at $\sim50\arcsec$. The disk is slightly
lopsided. We mask the dust lanes on and around the bulge.

NGC 7070: The galaxy has a distinctly blue nucleus of unknown nature and a
flattened bulge. We model the broken disk with a \sersic{} function. The disk is
slightly lopsided.

NGC 7079: The galaxy has a bar and shows a disk break at $\sim40\arcsec$.

NGC 7083: The disk has a Type~\Rmnum{2} profile and bears three major spiral
arms. It is part of the training sample presented in \citet{2017ApJ+Gao}. Here
we show the decomposition results that include all the above features (Model3 in
their Table~5). Note that the uncertainties are different from those presented
in their Table~5, since we include the model-induced uncertainties in this
study.

NGC 7098: The galaxy probably has a nuclear ring. The bar is embedded in a
lens/ring component.

NGC 7140: The galaxy has a nuclear ring. We model the disk break at the bar
radius but ignore the outer disk break at the edge of the spiral arms, treating
it as an outer feature when estimating bulge errors.  We mask the dust lanes on
and around the bulge and along the bar.

NGC 7144: The galaxy is classified as an elliptical in both HyperLeda and RC3,
but is recognized as an S0 in \citet{2013ApJ+Huang1}. It has a nuclear lens and
an inner lens.

NGC 7172: The galaxy is reported to host a Seyfert~2 nucleus
\citep{2010A&A+Veron-Cetty}, but we find no sign of the nucleus on the image,
probably due to severe dust attenuation. We mask the prominent dust lane across
the galaxy. We use two components to model the tidally distorted disk
\citep{1997ApJS+Turner}.

NGC 7192: The galaxy is classified as an elliptical in both HyperLeda and RC3,
but is recognized as an S0 in \citet{2013ApJ+Huang1}. It has a nuclear lens and
an inner lens.

NGC 7213: The galaxy has a Seyfert~1/LINER nucleus \citep{1979ApJ+Phillips,
  1984ApJ+Filippenko} and a nuclear ring. We mask the dust lanes around the
bulge. Significant sky gradient is present in the residuals.

NGC 7218: We model the disk break at $\sim20\arcsec$. We strive to mask the dust
lanes at the galaxy center. We find positive residuals of unknown physical nature
at the center, probably due to mismatch between the best-fit model and the data
caused by dust attenuation.

NGC 7314: The galaxy has a Seyfert~2 \citep{2014MNRAS+Asmus,2014A&A+Koulouridis}
or Seyfert~1 nucleus \citep{1986A&AS+Veron-Cetty,2010A&A+Veron-Cetty}. We model
the disk break at $\sim60\arcsec$ and mask the dust lanes near the bulge.

NGC 7329: The galaxy has a prominent bulge and bar. The disk shows grand-design
spiral arms that start from the inner ring. It is part of the training sample
presented in \citet{2017ApJ+Gao}. Here we show the decomposition results that
include all the above features (Model3 in their Table~10). Note that the
uncertainties are different from those presented in their Table~10, since we
include the model-induced uncertainties in this study.

NGC 7371: The galaxy has a weak and short bar embedded in a lens/ring
structure. We model the disk break at $\sim30\arcsec$. We ignore the extra light
of unknown physical nature in the galaxy outskirts ($\ga80\arcsec$) and treat it
as an outer feature when estimating bulge errors.

NGC 7377: The galaxy has a nuclear lens and an inner lens. The dust lanes are
masked during the fitting.

NGC 7392: The galaxy has a distinctly blue nucleus that was classified to be
inactive \citep{2003ApJS+Martini}. We fix $\alpha$ and $\beta$ of the weak bar
during the fitting. We model the disk break at the bar radius and at
$\sim70\arcsec$. We mask the spiral dust lanes approaching the center.

NGC 7412: We model the disk break at $\sim40\arcsec$. An extra disk component is
needed to account for the bar-like pattern produced by the spiral arms winding
onto the bulge. We mask the circumnuclear dust lanes during the fitting.

NGC 7418: The galaxy hosts an NSC \citep{2002AJ+Boker}, a weak bulge, and a weak
bar. We model the disk break at the edge of the spiral arms. The disk is
slightly lopsided. We mask the dust lanes around the bulge.

NGC 7421: The galaxy has an NSC \citep{2014MNRAS+Georgiev}. We model the disk
break at the bar radius. The disk is significantly lopsided.

NGC 7424: The galaxy has an NSC \citep{2002AJ+Boker}. The blue and flattened
bulge is embedded in a short and weak bar whose $\alpha$ and $\beta$ are fixed
during the fitting. We model the disk break at the bar radius but ignore the
outer disk break at the edge of the spiral arms, treating it as an outer feature
when estimating bulge errors. We mask the dust lanes on the bar.

NGC 7496: The galaxy has a distinctly blue nucleus with ambiguous
classifications: star-forming nucleus \citep{2010ApJ+Yuan} and Seyfert~2 nucleus
\citep{2010A&A+Veron-Cetty}. We model the disk break at the bar radius but
ignore the outer disk break at the edge of the spiral arms, treating it as an
outer feature when estimating bulge errors. We mask the dust lanes on the bulge
and along the bar.

NGC 7513: The galaxy has an NSC \citep{2002AJ+Carollo}. We fix $\alpha$ and
$\beta$ of the bar component, or else the bar will be unrealistically long. We
model the disk break at the bar radius but ignore the outer disk break at the
edge of the spiral arms, treating it as an outer feature when estimating bulge
errors. We mask the dust lanes on the bulge and along the bar.

NGC 7531: The galaxy has a distinctly red nucleus that was classified to be
Seyfert-like \citep{1986A&AS+Veron-Cetty}. We model both the inner lens and the
outer lens. We mask the dust lane running through the bulge.

NGC 7552: The bulge is embedded in the bar and lens. We ignore the disk break at
the edge of the spiral arms ($\sim100\arcsec$) and treat it as an outer feature
when estimating bulge errors.

NGC 7582: The galaxy is reported to host a Seyfert~2 nucleus
\citep{2010A&A+Veron-Cetty}, a star-forming nucleus \citep{2010ApJ+Yuan}, or a
composite nucleus \citep{1986A&AS+Veron-Cetty}. However, we find no sign of a
nucleus on the image, and attempts to include a PSF component in the fit
fails. The bar is embedded in a lens. We fix $\alpha$ and $\beta$ of the bar
component, or else the bar will be unrealistically long. We model the disk break
at the edge of the spiral arms.

NGC 7590: The galaxy is reported to host a Seyfert~2 nucleus
\citep{2010A&A+Veron-Cetty}. We find no sign of the nucleus on the
image. Forcibly including a PSF component does not impact the bulge parameters
much, and we simply ignore the purported nucleus. We model the disk break at the
edge of the spiral arms. We mask the dust lanes on and near the bulge.

NGC 7606: We model the disk break at the edge of spiral arms.

NGC 7689: We model the disk break at the edge of spiral arms.

NGC 7723: We model the disk break at the bar radius but ignore the outer disk
break at the edge of the spiral arms ($\sim80\arcsec$), treating it as an outer
feature when estimating bulge errors. We mask the dust lanes along the bar.

NGC 7755: The galaxy has a nuclear ring. We model the disk break at the bar
radius but ignore the outer disk break at the edge of the spiral arms, treating
it as an outer feature when estimating bulge errors. We mask the circumnuclear
dust lanes.